\documentclass[11pt]{article} 

\usepackage{fullpage} 
\usepackage{graphicx}
\usepackage{subcaption}
\usepackage{upref}
\usepackage{enumerate}
\usepackage{latexsym}
\usepackage{color,graphics}
\usepackage{relsize}
\usepackage{float}
\restylefloat{table}
\usepackage{algorithm}
\usepackage[redeflists]{IEEEtrantools}

\usepackage{amsmath,amsthm,amssymb,algorithmic,epsfig,graphicx, tikz}

\usepackage{amsfonts}
\usepackage{varioref}
\usepackage[ansinew]{inputenc}

\usepackage{qcircuit}

\usepackage{amsmath}
\usepackage{amsfonts}
\usepackage{tikz} 
\usepackage{multirow}

\usepackage{amssymb}
\usepackage{varioref}
\usepackage[ansinew]{inputenc}
\usepackage{authblk}

\newcommand{\norm}[1]{\left\lVert#1\right\rVert}

\newtheorem{theorem}{Theorem}[section]
\newtheorem{lemma}[theorem]{Lemma}
\newtheorem{claim}[theorem]{Claim}

\newtheorem{result}{Result}
\newtheorem{assumption*}{Assumption}
\newtheorem{definition}{Definition}

\newcommand{\R}{\mathbb{R}}
\newcommand{\Z}{\mathbb{Z}}

\def\bra#1{\mathinner{\langle{#1}|}}
\def\ket#1{\mathinner{|{#1}\rangle}}
\newcommand{\braket}[2]{\langle #1|#2\rangle}

\renewcommand{\part}[2]{\frac{\partial #1}{\partial #2}}

\newcommand{\als}[1]{\begin{align*}#1\end{align*}}

\newcommand{\en}[1]{\left ( #1 \right )}

\newcommand{\errdist}{\epsilon_1}
\newcommand{\errmult}{\epsilon_2}
\newcommand{\errnorms}{\epsilon_3}
\newcommand{\errtom}{\epsilon_4}
\newcommand{\errkappa}{\epsilon_{\tau}}
\newcommand{\n}{N}

\newcommand{\thmref}[1]{\hyperref[#1]{{Theorem~\ref*{#1}}}}
\newcommand{\lemref}[1]{\hyperref[#1]{{Lemma~\ref*{#1}}}}
\newcommand{\remref}[1]{\hyperref[#1]{{Remark~\ref*{#1}}}}
\newcommand{\corref}[1]{\hyperref[#1]{{Corollary~\ref*{#1}}}}
\newcommand{\eqnref}[1]{\hyperref[#1]{{Equation~(\ref*{#1})}}}
\newcommand{\claimref}[1]{\hyperref[#1]{{Claim~\ref*{#1}}}}
\newcommand{\remarkref}[1]{\hyperref[#1]{{Remark~\ref*{#1}}}}
\newcommand{\propref}[1]{\hyperref[#1]{{Proposition~\ref*{#1}}}}
\newcommand{\factref}[1]{\hyperref[#1]{{Fact~\ref*{#1}}}}
\newcommand{\defref}[1]{\hyperref[#1]{{Definition~\ref*{#1}}}}
\newcommand{\exampleref}[1]{\hyperref[#1]{{Example~\ref*{#1}}}}
\newcommand{\hypref}[1]{\hyperref[#1]{{Hypothesis~\ref*{#1}}}}
\newcommand{\secref}[1]{\hyperref[#1]{{Section~\ref*{#1}}}}
\newcommand{\chapref}[1]{\hyperref[#1]{{Chapter~\ref*{#1}}}}
\newcommand{\apref}[1]{\hyperref[#1]{{Appendix~\ref*{#1}}}}

\newcommand\blfootnote[1]{
  \begingroup
  \renewcommand\thefootnote{}\footnote{#1}
  \addtocounter{footnote}{-1}
  \endgroup
}

\AtBeginDocument{}

\begin{document}
\bstctlcite{IEEEexample:BSTcontrol}

\title{q-means: A quantum algorithm for unsupervised machine learning} 

%\author{ 
%Iordanis Kerenidis \thanks{
%CNRS, IRIF, Universit\'e Paris Diderot, Paris, France and  
%Centre for Quantum Technologies, National University of Singapore, Singapore. 
%Email: {\tt jkeren@irif.fr}.} 
%\and
%Jonas Landman \thanks{ IRIF, Universit\'e Paris Diderot, Paris, France. Ecole Polytechnique, Palaiseau, France.
%Email: {\tt jonas.landman@polytechnique.edu}.}   
%\and 
%Alessandro Luongo \thanks{Atos Bull, Les Clayes Sous Bois, France. IRIF, Universit\'e Paris Diderot, Paris, France Email: {\tt aluongo@irif.fr}.} 
%\and 
%Anupam Prakash \thanks{CNRS, IRIF, Universit\'e Paris Diderot, Paris, France Email: { \tt anupamprakash1@gmail.com}.}
%}

\author[1,2]{Iordanis Kerenidis}
\author[1,3]{Jonas Landman}
\author[1,4]{Alessandro Luongo}
\author[1]{Anupam Prakash}

\affil[1]{CNRS, IRIF, Universit\'e Paris Diderot, Paris, France}
\affil[2]{Centre for Quantum Technologies, National University of Singapore, Singapore}
\affil[3]{Ecole Polytechnique, Palaiseau, France}
\affil[4]{Atos Bull, Les Clayes Sous Bois, France}

\maketitle

\begin{abstract}
Quantum machine learning is one of the most promising applications of a full-scale quantum computer. 
Over the past few years, many quantum machine learning algorithms have been proposed that can potentially offer considerable speedups over the corresponding classical algorithms. 
In this paper, we introduce q-means, a new quantum algorithm for clustering which is  a canonical problem in unsupervised machine learning. The $q$-means algorithm has convergence and precision guarantees similar to $k$-means, and it outputs with high probability a good approximation of the $k$ cluster centroids like the classical algorithm. Given a dataset of $N$ $d$-dimensional vectors $v_i$ (seen as a matrix $V \in \mathbb{R}^{\n \times d})$ stored in QRAM, the running time of q-means is 
$\widetilde{O}\left(    k d \frac{\eta}{\delta^2}\kappa(V)(\mu(V) +  k \frac{\eta}{\delta}) + k^2 \frac{\eta^{1.5}}{\delta^2} \kappa(V)\mu(V)
\right)$ per iteration,  
where $\kappa(V)$ is the condition number, $\mu(V)$ is a parameter that appears in quantum linear algebra procedures and $\eta = \max_{i} \norm{v_{i}}^{2}$. 
For a natural notion of well-clusterable datasets, the running time becomes
$\widetilde{O}\left( k^2 d \frac{\eta^{2.5}}{\delta^3} + k^{2.5} \frac{\eta^2}{\delta^3} \right)$ per iteration, which is linear in the number of features $d$, and polynomial in the rank $k$, the maximum square norm $\eta$ and the error parameter $\delta$. Both running times are only polylogarithmic in the number of datapoints $N$.
Our algorithm provides substantial savings compared to the classical $k$-means algorithm that runs in time $O(kdN)$ per iteration, particularly for the case of large datasets.
\end{abstract} 

\blfootnote{Emails: jkeren@irif.fr, jonas.landman@polytechnique.edu, aluongo@irif.fr, anupam@irif.fr}

\thispagestyle{empty}

\newpage

\setcounter{page}{1}

\section{Introduction}
The last decade has witnessed the emergence of a scientific and industrial revolution, which leveraged our ability to process an increasing volume of data and extract value from it. Henceforth, the imminent widespread adoption of technologies such as the Internet of Things, IPv6, and 5G internet communications is expected to generate an even bigger amount of data, most of which will  be unlabelled. As the amount of data generated in our society is expected to grow, more powerful ways of processing information are needed. Quantum computation is a promising new paradigm for performing fast computations. In recent years, there have been proposals for quantum machine learning algorithms that have the potential to offer considerable  speedups over the corresponding classical algorithms, either exponential or large polynomial speedups \cite{ lloyd2014quantum, KP16, KP17, CGJ18, LMR13, kerenidis2018neural}. 

In most of these quantum machine learning applications, there are some common algorithmic primitives that are used to build the algorithms. For instance, quantum procedures for linear algebra (matrix multiplication,  inversion, and  projections in sub-eigenspaces of matrices), have been used for recommendation systems or dimensionality reduction techniques \cite{KP16, KL18, lloyd2014quantum}. Second, the ability to estimate distances between quantum states, for example through the SWAP test, has been used for supervised or unsupervised learning \cite{LMR13, WKS14}. We note that most of these procedures need quantum access to the data, which can be achieved by storing the data in specific data structures in a QRAM (Quantum Random Access Memory). 

Here, we are interested in unsupervised learning and in particular in the canonical problem of clustering: given a dataset represented as $\n$ vectors, we want to find an assignment of the vectors to one of $k$ labels (for a given $k$ that we assume to know) such that similar vectors are assigned to the same cluster. Often, the Euclidean distance is used to measure the similarity of vectors, but other metrics might be used, according to the problem under consideration.  

We propose \emph{$q$-means}, a quantum algorithm for \emph{clustering}, which can be viewed as a quantum alternative to the classical $k$-means algorithm. More precisely, $q$-means is the equivalent of the $\delta$-$k$-means algorithm, a robust version of $k$-means that will be defined later.  We provide a detailed analysis to show that $q$-means has an output consistent with the classical $\delta$-$k$-means algorithm and further has a running time that depends poly-logarithmically on the number of elements in the dataset. The last part of this work includes simulations which assert the performance and running time of the $q$-means algorithm. 

\subsection{Related Work}
In this section, we discuss previous work on quantum unsupervised learning and clustering. 
Aimeur, Brassard and Gambs \cite{aimeur2013quantum} gave two quantum algorithms for unsupervised learning using the amplification techniques from \cite{durr1996quantum}. Specifically, they proposed an algorithm for clustering based on minimum spanning trees that runs in time $\Theta(\n^{3/2})$ and a quantum algorithm for $k$-median (a problem related to k-means)  algorithm with complexity time $O(N^{3/2}/\sqrt{k})$.

Lloyd, Mohseni and Rebentrost \cite{LMR13} proposed quantum $k$-means and nearest centroid algorithms using an efficient subroutine for quantum distance estimation assuming as we do quantum access to the data. Given a dataset of $\n$ vectors in a feature space of dimension $d$, the running time of each iteration of the clustering algorithm (using a distance estimation procedure with error $\epsilon$) is  $O(\frac{k\n\log d}{\epsilon})$ to produce the quantum state corresponding to the clusters. Note that the time is linear in the number of data points and it will be linear in the dimension of the vectors if the algorithm needs to output the classical description of the clusters. 
% and of course. 
% the running time has to be multiplied by a factor of $d$ if a classical solution needs to be returned, which is necessary when a new iteration of the $k$-means algorithm needs to be executed. \iordanis{check above}

In the same work, they also proposed an adiabatic algorithm for the assignment step of the $k$-means algorithm, that can potentially provide an exponential speedup in the number of data points as well, in the case the adiabatic algorithm performs exponentially better than the classical algorithm. The adiabatic algorithm is used in two places for this algorithm, the first to select the initial centroids, and the second to assign data points to the closest cluster. However, while arguments are given for its efficiency, it is left as an open problem to determine how well the adiabatic algorithm performs on average, both in terms of the quality of solution and the running time.  

Wiebe, Kapoor and Svore \cite{WKS14} apply the minimum finding algorithm \cite{durr1996quantum} to obtain nearest-neighbor methods for supervised and unsupervised learning. At a high level, they recovered a Grover-type quadratic speedup with respect to the number of elements in the dataset in finding the $k$ nearest neighbors of a vector. 
Otterbach et al.  \cite{Otterbach17} performed clustering by exploiting a well-known reduction from clustering to the Maximum-Cut (MAXCUT) problem; the MAXCUT is then solved using QAOA, a quantum algorithm for performing approximate combinatorial optimization \cite{farhi2014quantum}.

Let us remark on a recent breakthrough by Tang et al. \cite{gilyen2018quantum, tang2018quantum, tang2018quantum2}, who proposed three classical machine learning algorithms obtained by dequantizing 
recommendation systems \cite{KP16} and low rank linear system solvers. Like the quantum algorithms, the running time of these classical algorithms is $O(\text{poly}(k) \text{polylog}(mn))$, that is poly-logarithmic in the dimension of the dataset and polynomial in the rank.  However, the polynomial dependence on the rank of the matrices is significantly worse than the quantum algorithms and in fact renders these classical algorithms highly impractical. For example, the new classical algorithm for stochastic regression inspired by the HHL algorithm \cite{harrow2009quantum} has a running time of $\tilde{O}(\kappa^{6} k^{16} \norm{A}_F^6 / \epsilon^6)$, which is impractical even for a rank-10 matrix. 

The extremely high dependence on the rank and the other parameters implies not only that the quantum algorithms are substantially faster (their dependence on the rank is sublinear!), but also that in practice there exist much faster classical algorithms for these problems. While the results of Tang et al. are based on the FKV methods \cite{frieze2004fast}, in classical linear algebra, algorithms based on the CUR decomposition that have a running time linear in the dimension and quadratic in the rank are preferred to the FKV methods \cite{frieze2004fast, drineas2004clustering, achlioptas2003fast}. For example, for the recommendation systems matrix of Amazon or Netflix, the dimension of the matrix is $10^6 \times 10^7$, while the rank is certainly not lower than $100$. The dependence on the dimension and rank of the quantum algorithm in \cite{KP16} is $O(\sqrt{k}\log(mn)) \approx O(10^2)$, of the classical CUR-based algorithm is $O(mk^2) \approx O(10^{11})$, while of the Tang algorithm is $O(k^{16} log(mn)) \approx O(10^{33})$.

It remains an open question to find classical algorithms for these machine learning problems that are poly-logarithmic in the dimension and are competitive with respect to the quantum or the classical algorithms for the same problems. This would involve using significantly different techniques than the ones presently used for these algorithms.

\subsection{The $k$-means algorithm}
The $k$-means algorithm was introduced in \cite{lloyd1982least}, and is extensively used for unsupervised problems. The inputs to $k$-means algorithm are vectors $v_{i} \in \R^{d}$ for $i \in [\n]$. These points must be partitioned in $k$ subsets according to a similarity measure,  which in k-means is the Euclidean distance between points. The output of the $k$-means algorithm is a list of $k$ cluster centers, which are called \textit{centroids}. 

The algorithm starts by selecting $k$ initial centroids randomly or using efficient heuristics like the $k$-means++ \cite{arthur2007k}. It then alternates between two steps: (i) Each data point is assigned the label of the closest centroid. (ii) Each centroid is updated to be the average of the data points assigned to the corresponding cluster. These two steps are repeated until convergence, that is, until the change in the centroids during one iteration is sufficiently small. 

More precisely, we are given a dataset $V$ of vectors $v_{i} \in \R^{d}$ for $i \in [\n]$. At step $t$, we denote the $k$ clusters by the sets $C_j^t$ for $j \in [k]$, and each corresponding centroid by the vector $c_{j}^{t}$. At each iteration, the data points $v_i$ are assigned to a cluster $C_j^t$ such that $C_1^t \cup C_2^t \cdots \cup C_K^t = V$ and $C_i^t \cap C_l^t = \emptyset$ for $i \neq l$. Let $d(v_{i}, c_{j}^{t})$ be the Euclidean distance between vectors $v_{i}$ and $c_{j}^{t}$.  
The first step of the algorithm assigns each $v_{i}$ a label $\ell(v_{i})^t$ corresponding to the closest centroid, that is 
$$\ell(v_{i})^{t} = \text{argmin}_{j \in [k]}(d(v_{i}, c_{j}^{t})).$$ The centroids are then updated, $c_{j}^{t+1} = \frac{1}{ |C_{j}^t|} \sum_{i \in C_{j}^t } v_{i},$
so that the new centroid is the average of all points that have been assigned to the cluster in this iteration. We say that we have converged if for a small threshold $\tau$ we have
$$\frac{1}{k}\sum_{j=1}^{k}{d(c_{j}^{t},c_{j}^{t-1}}) \leqslant \tau. $$ 
The loss function that this algorithm aims to minimize is the RSS (residual sums of squares), the sum of the squared distances between points and the centroid of their cluster. 
$$ \text{RSS} := \sum_{j \in [k]}\sum_{i\in C_j} d(c_j, v_i)^2 $$
The RSS decreases at each iteration of the $k$-means algorithm, the algorithm therefore converges to a local minimum for the RSS. The number of iterations $T$ for convergence depends on the data and the number of clusters.  A single iteration has complexity of $O(k\n d)$ since the $\n$ vectors of dimension $d$ have to be compared to each of the $k$ centroids. 

From a computational complexity point of view, we recall that it is NP-hard to find a clustering that achieves the global  minimum for the RSS. There are classical clustering algorithms based on optimizing different loss functions, however the k-means algorithm uses the RSS as the objective function. 
The algorithm can be super-polynomial in the worst case (the number of iterations is $2^{\omega(\sqrt{\n})}$  \cite{arthur2006slow}), 
but the number of iterations is usually small in practice. The $k$-means algorithm with a suitable heuristic like $k$-means++ to initialize the centroids finds a clustering such that the 
value for the RSS objective function is within a multiplicative $O(\log \n)$ factor of the minimum value \cite{arthur2007k}.

\subsection{$\delta$-$k$-means}
We now consider a $\delta$-robust version of the $k$-means in which we introduce some noise. The noise affects the algorithms in both of the steps of k-means: label assignment and centroid estimation. 

 Let us describe the rules for the assignment step of $\delta$-$k$-means more precisely. Let $c^{*}_i$ be the closest centroid to the data point $v_i$. Then, the set of possible labels $L_{\delta}(v_i)$ for $v_i$ is 
defined as follows:
$$L_{\delta}(v_i)  =  \{c_p  : | d^2(c^*_i, v_i ) - d^2(c_p, v_i) | \leq \delta \: \}$$ 
The assignment rule selects arbitrarily a cluster label from the set $L_{\delta}(v_i)$.  

Second, we add $\delta/2$ noise during the calculation of the centroid. Let $\mathcal{C}_j^{t+1}$ be the set of points which have been labeled by $j$ in the previous step. For $\delta$-k-means we pick a centroid $c^{t+1}_j $ with the property

$$ \norm{ c^{t+1}_j - \frac{1}{|\mathcal{C}^{t+1}_j|}\sum_{v_i \in \mathcal{C}^{t+1}_j} v_i} < \frac{\delta}{2}. $$

One way to do this is to calculate the centroid exactly and then add some small Gaussian noise to the vector to obtain the robust version of the centroid. 

Let us add two remarks on the $\delta$-$k$-means. First, for a well-clusterable data set and for a small $\delta$, the number of vectors on the boundary that risk to be misclassified in each step, that is the vectors for which $|L_{\delta}(v_i)|>1$ is typically much smaller compared to the vectors that are close to a unique centroid. Second, we also increase by $\delta/2$  the convergence threshold from the $k$-means algorithm.
All in all, $\delta$-$k$-means is able to find a clustering that is robust when the data points and the centroids are perturbed with some noise of magnitude $O(\delta)$. 
As we will see in this work, $q$-means is the quantum equivalent of $\delta$-$k$-means.

\subsection{Our results}
We define and analyse a new quantum algorithm for clustering, the $q$-means algorithm, whose performance is similar to that of the classical $\delta$-$k$-means algorithm and whose running time provides substantial savings, especially for the case of large data sets. 

The $q$-means algorithm combines most of the advantages that quantum machine learning algorithms can offer for clustering. First, the running time is poly-logarithmic in the number of elements of the dataset and depends only linearly on the dimension of the feature space. Second, $q$-means returns explicit classical descriptions of the cluster centroids that are obtained by the $\delta$-$k$-means algorithm. As the algorithm outputs a classical description of the centroids, it is possible to use them in further (classical or quantum) 
classification algorithms.

Our $q$-means algorithm requires that the dataset is stored in a QRAM (Quantum Random Access Memory), which allows the algorithm to use efficient linear algebra routines that have been developed using QRAM data structures. Of course, our algorithm can also be used for clustering datasets for which the data points can be efficiently prepared even without a QRAM, for example if the data points are the outputs of quantum circuits.

We start by providing a worst case analysis of the running time of our algorithm, which depends on parameters of the data matrix, for example the condition number and the parameter $\mu$ that appears in the quantum linear algebra procedures. Note that with $\widetilde{O}$ we hide polylogaritmic factors.

\begin{result}
Given dataset $V \in \mathbb{R}^{\n \times d} $ stored in QRAM, the q-means algorithm outputs with high probability centroids $c_1, \cdots,  c_k$ that are consistent with an output of the $\delta$-$k$-means algorithm in time 
$\widetilde{O}\left(    k d \frac{\eta}{\delta^2}\kappa(V)(\mu(V) + k \frac{\eta}{\delta}) + k^2 \frac{\eta^{1.5}}{\delta^2} \kappa(V)\mu(V)
\right)$ per iteration,  
%$\widetilde{O}\left( \frac{k\eta \kappa(V)} {\delta^{2}} \en{  \mu(V)(d+ k\sqrt{\eta}) + \frac{dk\eta}{ \delta}  }  \right)$
%$\widetilde{O} \left(  \frac{k d \eta}{\delta^2} \kappa(V) \left( \mu(V) + \frac{k \eta}{\delta} \right) + \frac{k^{2} \eta^{1.5} \kappa(V) \mu(V)}{ \delta^2}  \right)$  per iteration
where $\kappa(V)$ is the condition number, $\mu(V)$ is a parameter that appears in quantum linear algebra procedures and $1\leq \norm{v_i}^2 \leq \eta$. 
\end{result}

When we say that the $q$-means output is consistent with the $\delta$-$k$-means, we mean that with high probability the clusters that the $q$-means algorithm outputs are also possible outputs of the $\delta$-$k$-means.

We go further in our analysis and study a well-motivated model for datasets that allows for good clustering. We call these datasets {\em well-clusterable}. One possible way to think of such datasets is the following: a dataset is well-clusterable when the $k$ clusters arise from picking $k$ well-separated vectors as their centroids, and then each point in the cluster is sampled from a Gaussian distribution with small variance centered on the centroid of the cluster. We provide a rigorous definition in following sections. For such well-clusterable datasets we can provide a tighter analysis of the running time and have the following result, whose formal version appears as Theorem \ref{main}. 

\begin{result}
Given a well-clusterable dataset $V \in \mathbb{R}^{\n \times d} $ stored in QRAM, the q-means algorithm outputs with high probability $k$ centroids $c_1, \cdots ,c_k$ that are consistent with the output of the $\delta$-$k$-means algorithm in time 
$\widetilde{O}\left( k^2 d \frac{\eta^{2.5}}{\delta^3} + k^{2.5} \frac{\eta^2}{\delta^3} \right)$
per iteration, where $1\leq \norm{v_i}^2 \leq \eta$. 
\end{result}

\noindent 
In order to assess the running time and performance of our algorithm we performed extensive simulations for different datasets. The running time of the $q$-means algorithm is linear in the dimension $d$, which is necessary when outputting a classical description of the centroids, and polynomial in the number of clusters $k$ which is typically a small constant. The main advantage of the $q$-means algorithm is that it provably  depends logarithmically on the number of points, which can in many cases provide a substantial speedup. The parameter $\delta$ (which plays the same role as in the $\delta$-$k$-means) is expected to be a large enough constant that depends on the data and the parameter $\eta$ is again expected to be a small constant for datasets whose data points have roughly the same norm. For example, for the MNIST dataset, $\eta$ can be less than $8$ and $\delta$ can be taken to be equal to $0.5$. In Section \ref{experiment_section} we present the results of our simulations. For different datasets we find parameters $\delta$ such that the number of iterations is practically the same as in the $k$-means, and the $\delta$-$k$-means algorithm converges to a clustering that achieves an accuracy similar to the $k$-means algorithm or in times better. We obtained these simulation results by simulating the operations executed by the quantum algorithm adding the appropriate errors in the procedures.

\section{Quantum preliminaries}
We assume a basic understanding of quantum computing, we recommend Nielsen and Chuang \cite{NC02} for an introduction to the subject. A vector state $\ket{v}$ for $v \in \mathbb{R}^d$ is defined as $\ket{v} = \frac{1}{\norm{v} } \sum_{m \in [d] } v_m \ket{m}$, where $\ket{m}$ represents $e_{m}$, the $m^{th}$ vector in the standard basis. The dataset is represented by a matrix $V \in \R^{\n \times d}$, i.e. each row is a vector $v_i \in \R^{d}$ for $i \in [\n]$ that represents a single data point. 
The cluster centers, called centroids, at time $t$ are stored in the matrix $C^t \in \R^{k \times d}$, such that the $j^{th}$ row $c_{j}^{t}$ for $j\in [k]$ represents the centroid of the cluster $\mathcal{C}_j^t$. 

We denote as $V_k$ the optimal rank $k$ approximation of $V$, that is
 $V_{k} = \sum_{i=0}^k \sigma_i u_i v_i^T$, where $u_i, v_i$ are the row and column singular vectors respectively and the sum is over the largest $k$ singular values $\sigma_{i}$.  We denote as $V_{\geq \tau}$ the matrix  $\sum_{i=0}^\ell \sigma_i u_i v_i^T$ where $\sigma_\ell$ is the smallest singular value which is greater than $ \tau$.

We will assume at some steps that these matrices and $V$ and $C^{t}$ are stored in suitable QRAM data structures which are described in \cite{KP16}. To prove our results, we are going to use the following  tools:

\begin{theorem}[Amplitude estimation \cite{BHMT00}]\label{amplitudeamplification}
	Given a quantum algorithm $$A:\ket{0} \to \sqrt{p}\ket{y,1} + \sqrt{1-p}\ket{G,0}$$ where $\ket{G}$ is some garbage state, then for any positive integer $P$, the amplitude estimation algorithm outputs $\tilde{p}$ $(0 \le \tilde p \le 1)$ such that
	$$
	|\tilde{p}-p|\le 2\pi \frac{\sqrt{p(1-p)}}{P}+\left(\frac{\pi}{P}\right)^2
	$$
	with probability at least $8/\pi^2$. It uses exactly $P$ iterations of the algorithm $A$. 
	If $p=0$ then $\tilde{p}=0$ with certainty, and if $p=1$ and $P$ is even, then $\tilde{p}=1$ with certainty.\label{thm:ampest}
\end{theorem}

\noindent In addition to amplitude estimation, we will make use of a tool developed in \cite{WKS14} to boost the probability of getting a good estimate 
for the distances required for the $q$-means algorithm. In high level, we take multiple copies of the estimator from the amplitude estimation procedure, compute the median, and reverse the circuit to get rid of the garbage. Here we provide a theorem with respect to time and not query complexity.

\begin{theorem}[Median Evaluation  \cite{WKS14}]\label{median}
	Let $\mathcal{U}$ be a unitary operation that maps 
	$$\mathcal{U}:\ket{0^{\otimes n}}\mapsto \sqrt{a}\ket{x,1}+\sqrt{1-a} \ket{G,0}$$
	for some $1/2 < a \le 1$ in time $T$. Then there exists a quantum algorithm that, for any $\Delta>0$ and for any $1/2<a_0 \le a$, produces a state $\ket{\Psi}$ such that $\|\ket{\Psi}-\ket{0^{\otimes nL}}\ket{x}\|\le \sqrt{2\Delta}$ for some integer $L$, in time  
	$$
	2T\left\lceil\frac{\ln(1/\Delta)}{2\left(|a_0|-\frac{1}{2} \right)^2}\right\rceil.
	$$
	\label{lem:median}
\end{theorem}

We also need some state preparation procedures. These subroutines are needed for encoding vectors in $v_{i} \in \R^{d}$ into quantum states $\ket{v_{i}}$. An efficient state preparation procedure is provided by the QRAM data structures. 
\begin{theorem}[QRAM data structure \cite{KP16}]\label{qram}
	Let $V \in \mathbb{R}^{\n \times d}$, there is a data structure to store the rows of $V$ such that, 
	\begin{enumerate}
		\item The time to insert, update or delete a single entry $v_{ij}$ is $O(\log^{2}(N))$. 	
		\item A quantum algorithm with access to the data structure can perform the following unitaries in time $T=O(\log^{2}N)$. 
		\begin{enumerate} 
			\item $\ket{i}\ket{0} \to \ket{i}\ket{v_{i}} $ for $i \in [\n]$. 
			\item $\ket{0} \to \sum_{i \in [N]} \norm{v_{i}}\ket{i}$. 
		\end{enumerate}
	\end{enumerate}
\end{theorem}

In our algorithm we will also use subroutines for quantum linear algebra. 
For a symmetric matrix $M \in \R^{d\times d}$ with spectral norm $\norm{M}=1$ stored in the QRAM, 
the running time of these algorithms depends linearly on the condition number $\kappa(M)$ of the matrix, that can be replaced by $\kappa_\tau(M)$, a condition threshold where we keep only the singular values bigger than $\tau$, and the parameter $\mu(M)$, a matrix dependent parameter defined as
$$\mu(M)=\min_{p\in [0,1]} (\norm{M}_{F}, \sqrt{s_{2p}(M)s_{(1-2p)}(M^{T})}),$$ 
for $s_{p}(M) = \max_{i \in [n]} \sum_{j \in [d]} M_{ij}^{p}$.
The different terms in the minimum in the definition of $\mu(M)$ correspond to different choices for the data structure for storing $M$, as detailed in \cite{KP17}.  Note that $\mu(M) \leq \norm{M}_{F} \leq \sqrt{d}$ as we have assumed that $\norm{M}=1$. The running time also depends logarithmically on the relative error $\epsilon$ of the final outcome state.  \cite{CGJ18, GLSW18}. 

\begin{theorem}[Quantum linear algebra \cite{CGJ18} ]\label{qla}  Let $M \in \mathbb{R}^{d \times d}$ such that $\norm{M}_2 =1$ and $x \in \mathbb{R}^d$. Let $\epsilon,\delta>0$. If $M$  
is stored in appropriate QRAM data structures and the time to prepare $\ket{x}$ is $T_{x}$, then there exist quantum algorithms that with probability at least $1-1/poly(d)$ return
    \begin{enumerate}
        \item A state $\ket{z}$ such that $\norm{ \ket{z} - \ket{Mx}} \leq \epsilon$ in time $\widetilde{O}((\kappa(M)\mu(M) + T_{x} \kappa(M)) \log(1/\epsilon))$.  
        \item A state $\ket{z}$ such that $\norm{\ket{z} - \ket{M^{-1}x}} \leq \epsilon$ in time $\widetilde{O}((\kappa(M)\mu(M) + T_{x} \kappa(M)) \log(1/\epsilon))$.
        \item Norm estimate $z \in (1 \pm \delta)\norm{Mx}$, with relative error $\delta$, in time $\widetilde{O}(T_{x} \frac{\kappa(M)\mu(M)}{\delta}\log(1/\epsilon))$.
    \end{enumerate}
    
\end{theorem}
The linear algebra procedures above can also be applied to any rectangular matrix $V \in \mathbb{R}^{\n \times d}$ by considering instead the symmetric matrix $ \overline{V} = \left ( \begin{matrix}
0  &V \\ 
V^{T} &0 \\
\end{matrix}  \right )$.

The final component needed for the $q$-means algorithm is a linear time algorithm for vector state tomography that will be used to recover classical information from the quantum states corresponding to the new centroids in each step. Given a unitary $U$ that produces a quantum state $\ket{x}$, by calling $O(d \log{d}/\epsilon^{2})$ times $U$, the tomography algorithm is able to reconstruct a vector $\widetilde{x}$ that approximates $\ket{x}$ such that $\norm{ \ket{ \widetilde{x}} - \ket{x} } \leq \epsilon$.

\begin{theorem}  [Vector state tomography \cite{KP18}] \label{thm:tom} Given access to unitary $U$ such that $U\ket{0} = \ket{x}$ and its controlled version in time $T(U)$, there is a tomography algorithm with time complexity $O(T(U) \frac{ d \log d  }{\epsilon^{2}})$ that produces unit vector $\widetilde{x} \in \R^{d}$ such that $\norm{\widetilde{x}  - x }_{2} \leq \epsilon$ with probability at least $(1-1/poly(d))$. 
\end{theorem}

\section{Modelling well-clusterable datasets}\label{datasetassumption}
In this section, we define a model for the dataset in order to provide a tight analysis on the running time of our clustering algorithm. 
Note that we do not need this assumption for our general $q$-means algorithm, but in this model we can provide tighter bounds for its running time. 
Without loss of generality we consider in the remaining of the paper that the dataset $V$ is normalized so that for all $i \in [N]$, we have $1 \leq \norm{v_{i}}$, and we define the parameter $\eta = \max_i{\norm{v_i}^2}$. We will also assume that the number $k$ is the ``right'' number of clusters, meaning that we assume each cluster has at least some $\Omega(N/k)$ data points.

We now introduce the notion of a \emph{well-clusterable} dataset. The definition aims to capture some properties that we can expect from datasets that can be clustered efficiently using a k-means algorithm. Our notion of a well-clusterable dataset shares some similarity with the assumptions made in\cite{drineas2002competitive}, but there are also some differences specific to the clustering problem.

\begin{definition}[Well-clusterable dataset]\label{wcdataset}
A data matrix $V \in \R^{\n\times d}$ with rows $v_{i} \in \R^{d}, i \in [N]$ is said to be well-clusterable if there exist constants $\xi, \beta>0$, $\lambda \in [0,1]$, $\eta \leq 1$, and cluster centroids $c_i$ for $i\in [k]$ such that:
\begin{enumerate}
    \item (separation of cluster centroids): $d(c_i, c_j) \geq \xi \quad \forall i,j \in[k]  $
    \item (proximity to cluster centroid): At least $\lambda \n$ points $v_i$ in the dataset satisfy $d(v_i, c_{l(v_i)}) \leq \beta$ where $c_{l(v_i)}$ is the centroid 
    nearest to $v_{i}$. 
 \item (Intra-cluster smaller than inter-cluster square distances): 
 The following inequality is satisfied $$4\sqrt{\eta} \sqrt{ \lambda \beta^{2} + (1-\lambda) 4\eta} \leq \xi^{2} - 2\sqrt{\eta} \beta.$$
 
\end{enumerate}
\end{definition}

Intuitively, the assumptions guarantee that most of the data can be easily assigned to one of $k$ clusters, since these points are close to the centroids, and the centroids are sufficiently far from each other. The exact inequality comes from the error analysis, but in spirit it says that $\xi^2$ should be bigger than a quantity that depends on $\beta$ and the maximum norm $\eta$.

 We now show that a well-clusterable dataset has a good rank-$k$ approximation where $k$ is the number of clusters. This result will later be used for giving tight upper bounds on the running time of the quantum algorithm for well-clusterable datasets. As we said, one can easily construct such datasets by picking $k$ well separated vectors to serve as cluster centers and then each point in the cluster is sampled from a Gaussian distribution with small variance centered on the centroid of the cluster. 

\begin{claim}\label{low-rank}
 Let $V_{k}$ be the optimal $k$-rank approximation for a well-clusterable data matrix $V$, then $\norm{V-V_{k}}_{F}^2  \leq (  \lambda \beta^2 +(1-\lambda)4\eta)\norm{V}^2_F$.
\end{claim}

\begin{proof}
Let $W \in \R^{\n\times d}$ be the matrix with row $w_{i} = c_{l(v_i)}$, where $c_{l(v_i)}$ is the centroid closest to $v_i$. 
The matrix $W$ has rank at most $k$ as it has exactly $k$ distinct rows. As $V_k$ is the optimal 
rank-$k$ approximation to $V$, we have $\norm{V-V_{k}}_{F}^{2} \leq \norm{V-W}_F^2$. It therefore suffices to upper bound $\norm{V-W}_F^2$. 
Using the fact that $V$ is  well-clusterable, we have
    $$ \norm{V-W}_F^2 = \sum_{ij} (v_{ij} - w_{ij})^2 = \sum_{i} d(v_i, c_{l(v_i)})^2 \leq \lambda \n \beta^2 + (1-\lambda)\n4\eta,$$
where we used Definition \ref{wcdataset} to say that for a $\lambda N$ fraction of the 
points $d(v_i, c_{l(v_i)})^2 \leq \beta^{2}$ and for the remaining points $d(v_i, c_{l(v_i)})^2 \leq 4\eta$.     
Also, as all $v_{i}$ have norm at least $1$ we have $N \leq \norm{V}_F$, implying that 
  $\norm{V-V_k}^{2} \leq  \norm{V-W}_F^2 \leq ( \lambda \beta^2 +(1-\lambda)4\eta)\norm{V}_F^2$.

\end{proof}

\noindent The running time of the quantum linear algebra routines for the data matrix $V$ in Theorem \ref{qla} depend on the parameters $\mu(V)$ 
and $\kappa(V)$. We establish bounds on both of these parameters using the fact that $V$ is well-clusterable
\begin{claim}\label{mu}
Let $V$ be a well-clusterable data matrix, then $\mu(V):= \frac{\norm{V}_{F}}{\norm{V}} = O(\sqrt{k})$. 
\end{claim}
\begin{proof} 
We show that when we rescale $V$ so that $\norm{V} = 1$, then we have $\norm{V}_{F}= O(\sqrt{k})$ for the rescaled matrix. 
From the triangle inequality we have that
$\norm{V}_F \leq \norm{V-V_k}_F + \norm{V_k}_F$.
Using the fact that $\norm{V_k}_F^2 = \sum_{i \in [k]} \sigma_i^2 \leq k$ and Claim \ref{low-rank}, we have, 
$$\norm{V}_F \leq \sqrt{ (\lambda\beta^2+(1-\lambda)4\eta)} \norm{V}_F + \sqrt{k}$$
Rearranging, we have that
$\norm{V}_F \leq \frac{\sqrt{k}}{1-\sqrt{(\lambda\beta^2+(1-\lambda)4\eta)}} = O(\sqrt{k})$.  
\end{proof} 
We next show that if we use a condition threshold $\kappa_\tau(V)$ instead of the true condition number $\kappa(V)$, that is we consider 
the matrix $V_{\geq \tau} = \sum_{\sigma_{i} \geq \tau} \sigma_{i} u_{i} v_{i}^{T}$ by discarding the smaller singular values $\sigma_{i} < \tau$, the resulting matrix remains close to the original one, i.e. we have that $\norm{V - V_{\geq \tau}}_F$ is bounded.  
 
\begin{claim}\label{kappa}
Let $V$ be a matrix with a rank-$k$ approximation given by $\norm{V- V_{k}}_{F}\leq \epsilon' \norm{V}_{F}$ and let $\tau = \frac{\errkappa}{\sqrt{k}}\norm{V}_F$, then $\norm{V- V_{\geq \tau}}_{F}\leq (\epsilon' + \errkappa) \norm{V}_{F}$. 
\end{claim}
\begin{proof}
Let $l$ be the smallest index such that $\sigma_{l} \geq \tau$, so that we have $\norm{V-V_{\geq \tau}}_F = \norm{V-V_l}_F$. 
We split the argument into two cases depending on whether $l$ is smaller or greater than $k$. 
\begin{itemize}
	\item If $l \geq k$ then $\norm{V-V_l}_F \leq \norm{V-V_k}_F \leq  \epsilon' \norm{V}_F$.
	\item If $l < k$ then, $\norm{V-V_l}_F \leq \norm{V-V_k}_F + \norm{V_k-V_l}_F \leq  \epsilon' \norm{V}_F + \sqrt{\sum_{i=l+1}^k \sigma_i^2}$. 
	
	As each $\sigma_{i} < \tau$ and the sum is over at most $k$ indices, we have the upper bound  $(\epsilon' + \errkappa) \norm{V}_F$. 	
\end{itemize}
\end{proof}

\noindent
The reason we defined the notion of well-clusterable dataset is to be able to provide some strong guarantees for the clustering of most points in the dataset. Note that the clustering problem in the worst case is NP-hard and we only expect to have good results for datasets that have some good property. Intuitively, we should only expect $k$-means to work when the dataset can actually be clusterd in $k$ clusters. We show next that for a well-clusterable dataset $V$, there is a constant $\delta$ that can be computed in terms of the parameters in Definition \ref{wcdataset} such that the $\delta$-$k$-means clusters correctly most of the data points. 

\begin{claim}\label{distcentroid} 
Let $V$ be a well-clusterable data matrix. Then, for at least $\lambda \n$ data points $v_i$, we have 
$$\min_{j\neq \ell(i)}(d^2(v_i,c_j)-d^2(v_i,c_{\ell(i)}))\geq \xi^2 - 2\sqrt{\eta}\beta$$
which implies that a $\delta$-$k$-means algorithm with any $\delta < \xi^2 - 2\sqrt{\eta}\beta$ will cluster these points correctly.
\end{claim}
\begin{proof}
    By Definition \ref{wcdataset}, we know that for a well-clusterable dataset $V$, 
    we have that $d(v_i, c_{l(v_i)}) \leq \beta$ for at least $\lambda \n$ data points and where $c_{l(v_{i})}$ is the centroid closest to $v_{i}$. 
    Further, the distance between each pair of the $k$ centroids satisfies the bounds $2\sqrt{\eta} \geq d(c_i, c_j) \geq \xi$. By the triangle inequality, we have $d(v_i,c_j) \geq d(c_j,c_{\ell(i)})-d(v_i,c_{\ell(i)})$. Squaring both sides of the inequality and rearranging, 
  
    $$d^2(v_i,c_j) - d^2(v_i,c_{\ell(i)})\geq d^2(c_j,c_{\ell(i)})  - 2d(c_j,c_{\ell(i)})d(v_i,c_{\ell(i)}))$$
Substituting the bounds on the distances implied by the well-clusterability assumption, we obtain $d^2(v_i,c_j)-d^2(v_i,c_{\ell(i)}) \geq \xi^2 - 2\sqrt{\eta} \beta$. This implies that as long as we pick $\delta <  \xi^2 - 2\sqrt{\eta}\beta$, these points are assigned to the correct cluster, since all other centroids are more than $\delta$ further away than the correct centroid. 
    
\end{proof}

\section{The $q$-means algorithm}

The $q$-means algorithm is given as Algorithm \ref{$q$-means-algo}. At a high level, it follows the same steps as the classical $k$-means algorithm, where we now use quantum subroutines for distance estimation, finding the minimum value among a set of elements, matrix multiplication for obtaining the new centroids as quantum states, and efficient tomography. First, we pick some random initial points, using some classical tchnique, for example $k$-means$++$ \cite{arthur2007k}. Then, in Steps 1 and 2 all data points are assigned to clusters, and in Steps 3 and 4 we update the centroids of the clusters. The process is repeated until convergence. 

\begin{algorithm} 
\caption{$q$-means.} \label{$q$-means-algo}
\begin{algorithmic}[1]
 
\REQUIRE  Data matrix $V \in \R^{\n \times d}$ stored in QRAM data structure. Precision parameters $\delta$ for $k$-means, error parameters
$\errdist$ for distance estimation, $\errmult$ and $\errnorms$ for matrix multiplication and $\errtom$ for tomography. 
\ENSURE Outputs vectors $c_{1}, c_{2}, \cdots, c_{k} \in \R^{d}$ that correspond to the centroids at the final step of the $\delta$-$k$-means algorithm.\\
\vspace{10pt} 
\STATE Select $k$ initial centroids $c_{1}^{0}, \cdots, c_{k}^{0}$ and store them in QRAM data structure. 
\STATE t=0
\REPEAT 
\STATE {\bf Step 1: Centroid Distance Estimation}\\
Perform the mapping (Theorem \ref{dist}) 
\begin{equation}\label{initialstate}
\frac{1}{\sqrt{N}}\sum_{i=1}^{\n}   \ket{i} \otimes_{j \in [k]} \ket{j}\ket{0} \mapsto \frac{1}{\sqrt{N}}\sum_{i=1}^{\n}  \ket{i} \otimes_{j \in [k]} \ket{j}\ket{\overline{d^2(v_{i}, c_{j}^{t})}}
\end{equation}
where $|\overline{d^2(v_{i}, c_{j}^{t})} -  d^2(v_{i}, c_{j}^{t}) | \leq \epsilon_{1}. $\\
\STATE {\bf Step 2: Cluster Assignment}\\
Find the minimum distance among $\{d^2(v_{i}, c_{j}^{t})\}_{j\in[k]}$ (Lemma $\ref{lem:minimum}$), then uncompute Step 1 to create the superposition of all points and their labels
\begin{equation}
\frac{1}{\sqrt{N}}\sum_{i=1}^{\n}  \ket{i} \otimes_{j \in [k]} \ket{j}\ket{\overline{d^2(v_{i}, c_{j}^{t})}}
 \mapsto \frac{1}{\sqrt{N}}\sum_{i=1}^{\n} \ket{i} \ket{ \ell^t(v_{i})}
\end{equation} 

\STATE {\bf Step 3: Centroid states creation} \\
{\bf 3.1} Measure the label register to obtain a state $\ket{\chi_{j}^t} = \frac{1}{ \sqrt{ |\mathcal{C}^t_{j}|} }  \sum_{i\in \mathcal{C}^t_j}\ket{i}$, with prob. $\frac{|\mathcal{C}^{t}_j|}{N} $ \\ %= O(\frac{1}{k})$.\\
{\bf 3.2} Perform matrix multiplication with matrix $V^T$ and vector  $\ket{\chi_{j}^t}$  to obtain the state $\ket{c_{j}^{t+1}}$ with error $\errmult$, along with an estimation of $\norm{c_{j}^{t+1}}$ with relative error $\errnorms$ (Theorem \ref{qla}). \\

\STATE {\bf Step 4: Centroid Update} \\
{\bf 4.1} Perform tomography for the states $\ket{c_{j}^{t+1}}$ 
with precision $\errtom$ using the operation from Steps 1-3 (Theorem \ref{thm:tom}) and get a classical estimate $\overline{c}_j^{t+1}$ for the new centroids such that $|c_j^{t+1} - \overline{c}_j^{t+1}| \leq \sqrt{\eta}(\errnorms+\errtom) = \epsilon_{centroids}$\\ %$\norm{\ket{\overline{c_j^{t+1}}}-\ket{c_j^{t+1}}} \leqslant \errtom$ and $| \norm{c_j} - \overline{\norm{c_j}} | \leq \errnorms\norm{c_j}$
{\bf 4.2} Update the QRAM data structure for the centroids with the new vectors $\overline{c}^{t+1}_0 \cdots \overline{c}^{t+1}_k$. 

\STATE t=t+1
\UNTIL convergence condition is satisfied. 

\end{algorithmic}
\end{algorithm}

\subsection{Step 1: Centroid distance estimation} 

The first step of the algorithm estimates the square distance between data points and clusters using a quantum procedure. This can be done using the Swap Test as in \cite{LMR13} and also using the Frobenius distance estimation procedure \cite{KL18}. 
Indeed, the subroutine presented in \cite{KL18} (originally used to calculate the average square distance between a point and all points in a cluster) can be adapted to calculate the square distance or inner product (with sign) between two vectors stored in the QRAM. The distance estimation becomes very efficient when we have quantum access to the vectors and the centroids as in Theorem \ref{qram}. That is, when we can query the state preparation oracles
$
\ket{i}\ket{0} \mapsto \ket{i}\ket{v_i}, $ and $\ket{j}\ket{0} \mapsto \ket{j}\ket{c_j}
$ in time $T=O(\log d)$, and we can also query the norm of the vectors. 

For $q$-means, we need to estimate distances or inner products between vectors which have different norms. At a high level, if we first estimate the inner between the quantum states $\ket{v_i}$ and $\ket{c_j}$ corresponding to the normalized vectors and then multiply our estimator by the product of the vector norms we will get an estimator for the inner product of the unnormalised vectors. A similar calculation works for the square distance instead of the inner product.
If we have an absolute error $\epsilon$ for the square distance estimation of the normalized vectors, then the final error is of the order of $\epsilon \norm{v_i} \norm{c_j}$.  

We present now the distance estimation theorem we need for the $q$-means algorithm and develop its proof in the next subsection.

\begin{theorem}[Centroid Distance estimation]\label{dist}
	Let a data matrix $V \in \mathbb{R}^{\n \times d}$ and a centroid matrix $C \in \mathbb{R}^{k \times d}$ be stored in QRAM, such that the following unitaries $
\ket{i}\ket{0} \mapsto \ket{i}\ket{v_i}, $ and $\ket{j}\ket{0} \mapsto \ket{j}\ket{c_j}
$ can be performed in time $O(\log (Nd))$ and the norms of the vectors are known.
For any $\Delta > 0$ and $\errdist>0$, there exists a quantum algorithm that performs the mapping
$$\frac{1}{\sqrt{N}}\sum_{i=1}^{\n}  \ket{i} \otimes_{j \in [k]} ( \ket{j}\ket{0}) \mapsto\ \frac{1}{\sqrt{N}}\sum_{i=1}^{\n}  \ket{i} \otimes_{j \in [k]}(\ket{j}\ket{\overline{d^2(v_i,c_j)}}),$$ 
where $|\overline{d^{2}(v_i,c_j)}-d^{2}(v_i,c_j)| \leqslant  \errdist$ with probability at least $1-2\Delta$, in time $\widetilde{O}\left(k \frac{ \eta}{ \errdist} \log(1/\Delta) \right)$ where $\eta=\max_{i}(\norm{v_i}^2)$. 
\end{theorem}

\subsection{Proof of Theorem \ref{dist}}

The theorem will follow easily from the following lemma which computes the square distance or inner product of two vectors.

\begin{lemma}[Distance / Inner Products Estimation]\label{th:squareddistances}
	Assume for a data matrix $V \in \mathbb{R}^{\n \times d}$ and a centroid matrix $C \in \mathbb{R}^{k \times d}$ that the following unitaries $
\ket{i}\ket{0} \mapsto \ket{i}\ket{v_i}, $ and $\ket{j}\ket{0} \mapsto \ket{j}\ket{c_j}
$ can be performed in time $T$ and the norms of the vectors are known. For any $\Delta > 0$ and $\errdist>0$, there exists a quantum algorithm that  computes 
\begin{eqnarray*}
\ket{i}\ket{j}\ket{0} & \mapsto &  \ket{i}\ket{j}\ket{\overline{d^2(v_i,c_j)}}, \mbox{ where } |\overline{d^{2}(v_i,c_j)}-d^{2}(v_i,c_j)| \leqslant  \errdist \mbox{ with probability at least } 1-2\Delta, \mbox{ or } \\
\ket{i}\ket{j}\ket{0} & \mapsto &  \ket{i}\ket{j}\ket{\overline{(v_i,c_j)}}, \mbox{ where } |\overline{(v_i,c_j)}-(v_i,c_j)| \leqslant  \errdist \mbox{ with probability at least } 1-2\Delta
\end{eqnarray*}
in time $\widetilde{O}\left(\frac{ \norm{v_i}\norm{c_j} T \log(1/\Delta)}{ \errdist}\right)$. 
\end{lemma}

\begin{proof}

Let us start by describing a procedure to estimate the square $\ell_2$ distance between the normalised vectors $\ket{v_i}$ and $\ket{c_j}$. We start with the initial state 
$$
\ket{\phi_{ij}} := \ket{i} \ket{j} \frac{1}{\sqrt{2}}(\ket{0}+	\ket{1})\ket{0}
$$ 

Then, we query the state preparation oracle controlled on the third register to perform the mappings 
$\ket{i}\ket{j}\ket{0}\ket{0} \mapsto \ket{i}\ket{j}\ket{0}\ket{v_i}$ and $\ket{i}\ket{j}\ket{1}\ket{0} \mapsto \ket{i}\ket{j}\ket{1}\ket{c_j}$. 
The state after this is given by,

$$
\frac{1}{\sqrt{2}}\left( \ket{i}\ket{j}\ket{0}\ket{v_i} + \ket{i}\ket{j}\ket{1}\ket{c_j}\right)
$$
Finally, we apply an Hadamard gate on the the third register to obtain, 
$$
\ket{i}\ket{j}\left(
\frac{1}{2}\ket{0}\left(\ket{v_i} + \ket{c_j}\right)
+ \frac{1}{2}\ket{1}\left(\ket{v_i} - \ket{c_j}\right)
\right)
$$
The probability of obtaining $\ket{1}$ when the third register is measured is,
$$
p_{ij} =  \frac{1}{4}(2 - 2\braket{v_i}{c_j}) =  \frac{1}{4} d^2(\ket{v_i}, \ket{c_j}) =  \frac{1 - \langle v_i | c_j\rangle}{2}
$$ 
which is proportional to the square distance between the two normalised vectors.

We can rewrite $\ket{1}\left(\ket{v_i} - \ket{c_j}\right)$ as $\ket{y_{ij},1}$ (by swapping the registers), and hence we have the final mapping
\begin{equation}\label{QDE}
A: \ket{i}\ket{j} \ket{0} \mapsto \ket{i}\ket{j}(\sqrt{p_{ij}}\ket{y_{ij},1}+\sqrt{1-p_{ij}}\ket{G_{ij},0}) 
\end{equation}
where the probability $p_{ij}$ is proportional to the square distance between the normalised vectors and $G_{ij}$ is a garbage state. Note that the running time of $A$ is $T_A=\tilde{O}(T)$.

Now that we know how to apply the transformation described in Equation \ref{QDE}, we can use known techniques to perform the centroid distance estimation as defined in Theorem \ref{dist} within additive error $\epsilon$ with high probability. The method uses two tools, amplitude estimation, and the median evaluation \ref{median} from \cite{WKS14}. 

First, using amplitude estimation (Theorem \ref{amplitudeamplification}) with the unitary $A$ defined in Equation \ref{QDE}, we can create a unitary operation that maps
$$
\mathcal{U}: \ket{i}\ket{j}  \ket{0} \mapsto \ket{i}\ket{j} \left( \sqrt{\alpha}  \ket{ \overline{p_{ij}}, G, 1} + \sqrt{ (1-\alpha ) }\ket{G', 0}  \right) 
$$ 
where $G, G'$ are garbage registers, $|\overline{p_{ij}} - p_{ij}  |  \leq \epsilon$ and $\alpha \geq 8/\pi^2$. 
The unitary $\mathcal{U}$ requires $P$ iterations of $A$ with $P=O(1/\epsilon)$. Amplitude estimation thus takes time $T_{\mathcal{U}} = \widetilde{O}(T/\epsilon)$. We can now apply Theorem \ref{median} for the unitary $\mathcal{U}$ to obtain a quantum state $\ket{\Psi_{ij}}$ such that, 

$$\|\ket{\Psi_{ij}}-\ket{0}^{\otimes L}\ket{\overline{p_{ij}}, G}\|_2\le \sqrt{2\Delta}$$

The running time of the procedure is 
$O( T_{\mathcal{U}} \ln(1/\Delta)) = \widetilde{O}( \frac{T }{\epsilon}\log (1/\Delta)  ) $.

Note that we can easily multiply the value $\overline{p_{ij}}$ by 4 in order to have the estimator of the square distance of the normalised vectors or compute $1-2\overline{p_{ij}}$ for the normalized inner product. Last, the garbage state does not cause any problem in calculating the minimum in the next step, after which this step is uncomputed. 

The running time of the procedure is thus
$O( T_{\mathcal{U}} \ln(1/\Delta)) = O( \frac{T }{\epsilon}\log (1/\Delta)  )$. 

The last step is to show how to estimate the square distance or the inner product of the unnormalised vectors. Since we know the norms of the vectors, we can simply multiply the estimator of the normalised inner product with the product of the two norms to get an estimate for the inner product of the unnormalised vectors and a similar calculation works for the distance. Note that the absolute error $\epsilon$ now becomes $\epsilon \norm{v_i}\norm{c_j}$ and hence if we want to have in the end an absolute error $\epsilon$ this will incur a factor of $\norm{v_i}\norm{c_j}$ in the running time. This concludes the proof of the lemma. 
\end{proof}

The proof of the theorem follows rather straightforwardly from this lemma. In fact one just needs to apply the above distance estimation procedure from Lemma \ref{th:squareddistances} $k$ times. Note also that the norms of the centroids are always smaller than the maximum norm of a data point which gives us the factor $\eta. $

\subsection{Step 2: Cluster assignment}\label{labels1} 
At the end of step 1, we have coherently estimated the square distance between each point in the dataset and the $k$ centroids in separate registers. We can now select the index $j$ that corresponds to the centroid closest to the given data point, written as $\ell(v_{i}) = \text{argmin}_{j \in [k]}(d(v_{i}, c_{j}))$. 
As the square is a monotone function, we do not need to compute the square root of the distance in order to find $\ell(v_{i})$. 
\begin{lemma}[Circuit for finding the minimum]\label{lem:minimum}
	Given $k$ different $\log p$-bit registers $\otimes_{j \in [k]} \ket{a_j}$, there is a quantum circuit $U_{min}$ that maps
	$(\otimes_{j \in [p]} \ket{a_j})\ket{0} \to (\otimes_{j \in [k]}\ket{a_j})\ket{\text{\em argmin}(a_j)}$ in time ${O}(k \log p)$. 
\end{lemma}
\begin{proof} 
We append an additional register for the result that is initialized to $\ket{1}$. We then repeat the following operation for $2\leq j \leq k$, we compare registers $1$ and $j$, if the value in register $j$ is smaller we swap registers $1$ and $j$ and update the result register to $j$. The cost of the procedure is ${O}(k \log p)$.   	
\end{proof}
\noindent The cost of finding the minimum is $\widetilde{O}(k)$ in step 2 of the $q$-means algorithm, while we also need to uncompute the distances by repeating Step 1. 
Once we apply the minimum finding Lemma $\ref{lem:minimum}$ and undo the computation we obtain the state
\begin{equation}\label{labels} 
\ket{\psi^t} := \frac{1}{\sqrt{N}}\sum_{i=1}^{\n} \ket{i} \ket{ \ell^t(v_{i})}.
\end{equation}

\subsection{Step 3: Centroid state creation} \label{create_centroids}
The previous step gave us the state $\ket{\psi^t}= \frac{1}{\sqrt{N}}\sum_{i=1}^{\n} \ket{i} \ket{ \ell^t(v_{i})}$.
The first register of this state stores the index of the data points while the second register stores the label for the data point in the current iteration. 
Given these states, we need to find the new centroids  $\ket{c_j^{t+1}}$, which are the average of the data points having the same label. 

Let $\chi_{j}^{t} \in \R^{N}$ be the characteristic vector for cluster $j \in [k]$ at iteration $t$ scaled to unit $\ell_{1}$ norm, that is $(\chi_{j}^{t})_{i} = \frac{1}{  |C_{j}^{t}|} $ if $i\in \mathcal{C}_{j}$ and $0$ if $i \not \in \mathcal{C}_{j}$. The creation of the quantum states corresponding to the centroids is based on the 
following simple claim. 
\begin{claim} \label{simple} 
Let $\chi_{j}^{t} \in \R^{N}$ be the scaled characteristic vector for $\mathcal{C}_{j}$ at iteration $t$ and $V \in\R^{\n\times d}$ be the data matrix, then $c_{j}^{t+1} = V^{T} \chi_{j}^{t}$. 
\end{claim} 
\begin{proof} 
The $k$-means update rule for the centroids is given by $c_{j}^{t+1} = \frac{1}{ |C_{j}^{t}|} \sum_{i \in C_{j} } v_{i}$. As the columns of $V^{T}$ are the vectors $v_{i}$, this can be rewritten as 
$c_{j}^{t+1} = V^{T} \chi_{j}^{t}$. 
\end{proof} 
\noindent The above claim allows us to compute the updated centroids $c_{j}^{t+1}$ using quantum linear algebra operations.
In fact, the state $\ket{\psi^t}$ can be written as a weighted superposition of the characteristic vectors of the clusters. 
\als{ 
\ket{\psi^t} = \sum_{j=1}^{k}\sqrt{\frac{|C_{j}|}{N}} \left( \frac{1}{ \sqrt{ |C_{j}|} }  \sum_{i\in \mathcal{C}_j}\ket{i}\right)\ket{j} = \sum_{j=1}^{k}\sqrt{\frac{|C_{j}|}{N}}
 \ket{\chi_{j}^{t} } \ket{j} 
 } 
 
By measuring the last register, we can sample from the states $\ket{\chi_{j}^{t} }$ for $j \in [k]$, with probability proportional to the size of the cluster. We assume here that all $k$ clusters are non-vanishing, in other words they have size $\Omega(N/k)$. 
Given the ability to create the states $\ket{\chi_{j}^{t} }$ and given that the matrix $V$ is stored in QRAM, we can now perform quantum matrix multiplication by $V^T$ to recover an approximation of the state $\ket{V^T\chi_{j}}=\ket{c_{j}^{t+1}}$ with error $\errmult$, as stated in Theorem \ref{qla}. Note that the error $\errmult$ only appears inside a logarithm. The same Theorem allows us to get an estimate of the norm $\norm{V^T\chi_{j}^{t}}=\norm{c_{j}^{t+1}}$ with relative error $\errnorms$. For this, we also need an estimate of the size of each cluster, namely the norms $\norm{\chi_{j}}$. We already have this, since the measurements of the last register give us this estimate, and since the number of measurements made is large compared to $k$ (they depend on $d$), the error from this source is negligible compared to other errors. 

 The running time of this step is derived from Theorem  
\ref{qla} where the time to prepare the state $\ket{\chi_{j}^{t}}$ is the time of Steps 1 and 2. Note that we do not have to add an extra $k$ factor due to the sampling, since we can run the matrix multiplication procedures in parallel for all $j$ so that every time we measure a random $\ket{\chi_{j}^{t}}$ we perform one more step of the corresponding matrix multiplication. Assuming that all clusters have size $\Omega(N/k)$ we will have an extra factor of $O(\log k)$ in the running time by a standard coupon collector argument.

\subsection{Step 4: Centroid update} \label{update_centroids}

In Step 4, we need to go from quantum states corresponding to the centroids, to a classical description of the centroids in order to perform the update step. For this, we will apply the vector state tomography algorithm, stated in Theorem \ref{thm:tom}, on the states $\ket{c_{j}^{t+1}}$ that we create in Step 3. 
Note that for each $j \in [k]$ we will need to invoke the unitary that creates the states $\ket{c_{j}^{t+1}}$ a total of $O(\frac{d \log d}{\errtom^{2}})$ times for achieving $\norm{\ket{c_j} - \ket{\overline{c_j}}} < \errtom$. Hence, for performing the tomography of all clusters, we will invoke the unitary $O(\frac{k (\log k) d(\log d)}{\errtom^{2}})$ times where the $O(k\log k)$ term is the time to get a copy of each centroid state. 

The vector state tomography gives us a classical estimate of the unit norm centroids within error $\errtom$, that is $\norm{\ket{c_j} - \ket{\overline{c_j}}} < \errtom$. Using the approximation of the norms $\norm{c_{j}}$ with relative error $\errnorms$ from Step 3, we can combine these estimates to recover the centroids as vectors. The analysis is described in the following claim:

\begin{claim}\label{epsiloncentroid}
Let $\errtom$ be the error we commit in estimating $\ket{c_j}$ such that $\norm{ \ket{c_j} - \ket{\overline{c_j}}} < \errtom$, and $\errnorms$ the error we commit in the estimating the norms,  $|\norm{c_j} - \overline{\norm{c_j}}| \leq \errnorms \norm{c_{j}} $. Then $\norm{\overline{c_j} - c_j} \leq \sqrt{\eta} (\errnorms +  \errtom) = \epsilon_{centroid}$. 
\end{claim}

\begin{proof}
We can rewrite $\norm{c_j - \overline{c_j}}$ as $\norm{ \norm{c_j}\ket{c_j} - \overline{\norm{c_j}}\ket{\overline{c_j}}}$. It follows from triangle inequality that:
$$ \norm{\overline{\norm{c_j}}\ket{\overline{c_j}} - \norm{c_j}\ket{c_j}}  \leq \norm{\overline{\norm{c_j}}\ket{\overline{c_j}} -  \norm{c_j}\ket{\overline{c_j}}} + \norm{\norm{c_j}\ket{\overline{c_j}} -  \norm{c_j}\ket{c_j}} $$
We have the upper bound $\norm{c_{j}} \leq \sqrt{\eta}$. 
Using the bounds for the error we have from tomography and norm estimation, we can upper bound the first term by $\sqrt{\eta} \errnorms$ and the second term by $\sqrt{\eta} \errtom$. The claim follows.  	
\end{proof}

Let us make a remark about the ability to use Theorem \ref{thm:tom} to perform tomography in our case.
The updated centroids will be recovered in step 4 using the vector state tomography algorithm in Theorem \ref{thm:tom} on the composition of the unitary that prepares $\ket{\psi^{t}}$ and the unitary that multiplies the first register of $\ket{\psi^t}$ by the matrix $V^{T}$. The input of the tomography algorithm requires a unitary $U$ 
such that $U\ket{0} = \ket{x}$ for a fixed quantum state $\ket{x}$. However, the labels $\ell(v_{i})$ are not deterministic due to errors in distance estimation, 
hence the composed unitary $U$ as defined above therefore does not produce a fixed pure state $\ket{x}$. 

We therefore need a procedure that finds labels $\ell(v_{i})$ that are a deterministic function of $v_{i}$ and the centroids $c_{j}$ for $j \in [k]$. One solution is to 
change the update rule of the $\delta$-$k$-means algorithm to the following: Let $\ell(v_{i})= j$ if $\overline{d(v_{i}, c_{j})}   < \overline{d(v_{i}, c_{j'})} - 2\delta$ for $j' \neq j$ where 
we discard the points to which no label can be assigned. This assignment rule ensures that if the second register is measured and found to be in state 
$\ket{j}$, then the first register contains a uniform superposition of points from cluster $j$ that are $\delta$ far from the cluster boundary (and possibly a few points 
that are $\delta$ close to the cluster boundary). Note that this simulates exactly the $\delta$-$k$-means update rule while discarding some of the data points 
close to the cluster boundary. The $k$-means centroids are robust under such perturbations, so we expect this assignment rule to produce 
good results in practice. 

A better solution is to use consistent phase estimation instead of the usual phase estimation for the distance estimation step , which can be found in \cite{T13, A12}. 
The distance estimates are generated by the phase estimation algorithm applied to a certain unitary in the amplitude estimation step. The usual phase estimation algorithm does not produce a deterministic answer and 
instead for each eigenvalue $\lambda$ outputs with high probability one of two possible estimates $\overline{\lambda}$ such that $|\lambda - \overline{\lambda}|\leq \epsilon$. 
Instead, here as in some other applications we need the consistent phase estimation algorithm that with high probability outputs a deterministic estimate such that $|\lambda - \overline{\lambda}|\leq \epsilon$. 

We also describe another simple method of getting such consistent phase estimation, which is to combine phase estimation estimates that are obtained for two different precision values. Let us assume that the 
eigenvalues for the unitary $U$ are $e^{2\pi i \theta_{i}}$ for $\theta_{i} \in [0, 1]$. First, we perform phase estimation with precision $\frac{1}{N_{1}}$ where $N_{1}=2^{l}$ is a power 
of $2$. We repeat this procedure $O(\log N/\theta^{2})$ times and output the median estimate. If the value being estimated 
is $\frac{\lambda + \alpha }{2^{l}}$ for $\lambda \in \Z$ and $\alpha \in [0,1]$ and $|\alpha - 1/2 | \geq \theta'$ for an explicit constant $\theta'$ (depending on $\theta$) then 
with probability at least $1-1/\text{poly}(N)$ the median estimate will be unique and will equal to $1/2^{l}$ times the closest integer to $(\lambda+ \alpha)$. 
In order to also produce a consistent estimate for the eigenvalues for the cases where the above procedure fails, we perform a second phase estimation with precision $2/3N_{1}$. 
We repeat this procedure as above for $O(\log N/\theta^{2})$ iterations and taking the median estimate. The second procedure fails to produce a consistent estimate only 
for eigenvalues $\frac{\lambda + \alpha }{2^{l}}$ for $\lambda \in \Z$ and $\alpha \in [0,1]$ and $|\alpha - 1/3 | \leq \theta'$ or $|\alpha - 2/3 | \leq \theta'$ for a suitable constant 
$\theta'$. Since the cases where the two procedures fail are mutually exclusive, one of them succeeds with probability $1-1/\text{poly}(N)$. The estimate produced 
by the phase estimation procedure is therefore deterministic with very high probability. In order to complete this proof sketch, we would have to give explicit values of the constants $\theta$ and $\theta'$ 
and the success probability, using the known distribution of outcomes for phase estimation. 

For what follows, we assume that indeed the state in Equation \ref{labels} is almost a  deterministic state, meaning that when we repeat the procedure we get the same state with very high probability.

We set the error on the matrix multiplication to be $\errmult \ll \frac{\errtom^2}{d\log d}$ as we need to call the unitary that builds $c^{t+1}_j$ for $O(\frac{d\log d}{\errtom^2})$ times. We will see that this does not increase the runtime of the algorithm, as the dependence of the runtime for matrix multiplication is logarithmic in the error.

\section{Analysis} \label{analysis}

We provide our general theorem about the running time and accuracy of the $q$-means algorithm.

\begin{theorem}[$q$-means]\label{$q$-meansgeneral}
For a data matrix $V \in \mathbb{R}^{\n \times d}$ stored in an appropriate QRAM data structure and parameter $\delta >0$, the q-means algorithm with high probability outputs centroids consistent with the classical $\delta$-$k$-means algorithm, in time 
$\widetilde{O}\left(    k d \frac{\eta}{\delta^2}\kappa(V)(\mu(V) + k \frac{\eta}{\delta}) + k^2 \frac{\eta^{1.5}}{\delta^2} \kappa(V)\mu(V)
\right)$ per iteration,  
%$\widetilde{O}\left( \frac{k\eta \kappa(V)} {\delta^{2}} \en{  \mu(V)(d+ k\sqrt{\eta}) + \frac{dk\eta}{ \delta}  }  \right)$ per iteration. 
%$\widetilde{O} \left(  \kappa(V) \mu(V) (\frac{\eta k d + k^{2} \eta^{1.5} }{ \delta^{2}} ) + \frac{\kappa(V)k^{2} \eta^{2} d} {\delta^{3}}\right)$
where $\kappa(V)$ is the condition number, 
$\mu(M)=\min_{p\in [0,1]} (\norm{M}_{F}, \sqrt{s_{2p}(M)s_{(1-2p)}(M^{T})}),$ and $1\leq \norm{v_i}^2 \leq \eta$. 
\end{theorem}

We prove the theorem in Sections \ref{erroranalysis} and \ref{runtimeanalysis} and then provide the running time of the algorithm for well-clusterable datasets as Theorem \ref{main}.

\subsection{Error analysis}\label{erroranalysis}
In this section we determine the error parameters in the different steps of the quantum algorithm so that the quantum algorithm behaves the same as the classical $\delta$-$k$-means. More precisely, we will determine the values of the errors $\errdist, \errmult, \errnorms,\errtom$ in terms of $\delta$ so that firstly, the cluster assignment of all data points made by the $q$-means algorithm is consistent with a classical run of the $\delta$-$k$-means algorithm, and also that the centroids computed by the $q$-means after each iteration are again consistent with centroids that can be returned by the $\delta$-$k$-means algorithm. 

The cluster assignment in $q$-means happens in two steps. The first step estimates the square distances between all points and all centroids. The error in this procedure is of the form
$$ |\overline{d^2(c_j,v_i)} - d^2(c_j,v_i) | < \errdist.$$
for a point $v_i$ and a centroid $c_j$.
The second step finds the minimum of these distances without adding any error.

For the $q$-means to output a cluster assignment consistent with the $\delta$-$k$-means algorithm, we require that: 
$$\forall j \in [k], \quad | \overline{d^2(c_j,v_i) } - d^2(c_j,v_i)  | \leq \frac{\delta}{2}$$
which implies that no centroid with distance more than $\delta$ above the minimum distance can be chosen by the $q$-means algorithm as the label. Thus we need to take 
$\errdist < \delta/2$. 

After the cluster assignment of the $q$-means (which happens in superposition), we update the clusters, by first performing a matrix multiplication to create the centroid states and estimate their norms, and then a tomography to get a classical description of the centroids.
The error in this part is $\epsilon_{centroids}$, as defined in Claim \ref{epsiloncentroid}, namely 

$$\norm{\overline{c}_{j} - c_j} \leq \epsilon_{centroid}  = \sqrt{\eta} (\errnorms + \errtom).$$ 

Again, for ensuring that the $q$-means is consistent with the classical $\delta$-$k$-means algorithm we take
$\errnorms < \frac{\delta}{4\sqrt{\eta}}$ and $\errtom < \frac{\delta}{4\sqrt{\eta}}$. Note also that we have ignored the error $\errmult$ that we can easily deal with since it only appears in a logarithmic factor.

\subsection{Runtime analysis}\label{runtimeanalysis}
As the classical algorithm, the runtime of $q$-means depends linearly on the number of iterations, so here we analyze the cost of a single step.

The cost of tomography for the $k$ centroid vectors is $O(\frac{kd \log k \log d}{{\errtom}^{2}})$ times the cost of preparation of a single centroid state $\ket{c_{j}^{t}}$. 
A single copy of $\ket{c_{j}^{t}}$ is prepared applying the matrix multiplication by $V^{T}$ procedure on the state $\ket{\chi_{j}^{t}}$ obtained using square distance estimation. The time required for preparing a single copy of $\ket{c_{j}^{t}}$ is $O( \kappa(V) (\mu(V) + T_{\chi}) \log (1/\epsilon_{2}))$ by  Theorem \ref{qla} where $T_{\chi}$ is the time for preparing $\ket{\chi_{j}^{t}}$.  The time $T_{\chi}$  is $\widetilde{O}\left(\frac{k\eta\log(\Delta^{-1})\log(\n d)}{ \errdist}\right)= \widetilde{O}(\frac{k\eta} { \errdist} )$ by Theorem \ref{dist}. 

The cost of norm estimation for $k$ different centroids is independent of the tomography cost and is $\widetilde{O}( \frac{k T_{\chi} \kappa(V) \mu(V) }{\epsilon_{3}} )$. 
Combining together all these costs and suppressing all the logarithmic factors we have a total running time of, 

\als{ 
\widetilde{O} \en{ kd \frac{1 }{ \errtom^{2}}  \kappa(V) \en{  \mu(V) +  k \frac{\eta} { \errdist} } + k^{2} \frac{ \eta }{\epsilon_{3} \epsilon_{1} } \kappa(V) \mu(V) } 
} 
The analysis in section \ref{erroranalysis} shows that we can take $\errdist = \delta/2$, $\errnorms = \frac{\delta}{4\sqrt{\eta}}$ and $\errtom = \frac{\delta}{4\sqrt{\eta}}$. 
Substituting these values in the above running time, it follows that the running time of the $q$-means algorithm is 
$$\widetilde{O} \left(  k d  \frac{\eta}{\delta^2} \kappa(V) \left( \mu(V) + k \frac{ \eta}{\delta} \right) + k^{2}\frac{ \eta^{1.5} }{ \delta^2} \kappa(V) \mu(V) \right).$$

%= \widetilde{O}\left( \frac{k\eta \kappa(V)} {\delta^{2}} \en{  \mu(V)(d+ k\sqrt{\eta}) + \frac{dk\eta}{ \delta}  }  \right) $.

\noindent This completes the proof of Theorem \ref{$q$-meansgeneral}. 
We next state our main result when applied to a well-clusterable dataset, as in Definition \ref{datasetassumption}.

\begin{theorem}[$q$-means on well-clusterable data]\label{main}
For a well-clusterable dataset $V \in \mathbb{R}^{\n \times d}$ stored in appropriate QRAM, the q-means algorithm returns with high probability the $k$ centroids consistently with the classical $\delta$-$k$-means algorithm for a constant $\delta$ in time 
$\widetilde{O}\left( k^2 d \frac{\eta^{2.5}}{\delta^3} + k^{2.5} \frac{\eta^2}{\delta^3} \right)$ per iteration, for $1\leq \norm{v_i}^2 \leq \eta$.  
\end{theorem}

\begin{proof}
Let $V \in \mathbb{R}^{\n \times d}$ be a well-clusterable dataset as in Definition \ref{wcdataset}. In this case, we know by Claim \ref{kappa} that $\kappa(V)=\frac{1}{\sigma_{min}}$ can be replaced by a thresholded condition number $\kappa_\tau(V)=\frac{1}{\tau}$. In practice, this is done by discarding the singular values smaller than a certain threshold during quantum matrix multiplication. Remember that by Claim \ref{mu}  we know that $\norm{V}_F = O(\sqrt{k})$. Therefore we need to pick $\epsilon_\tau$ for a threshold $\tau = \frac{\errkappa}{\sqrt{k}}\norm{V}_F$ such that $\kappa_\tau(V) = O(\frac{1}{\epsilon_{\tau}})$.   

Thresholding the singular values in the matrix multiplication step introduces an additional additive error in $\epsilon_{centroid}$. By Claim \ref{kappa} and Claim \ref{epsiloncentroid} , we have that the error $\epsilon_{centroid}$ in approximating the true centroids becomes $\sqrt{\eta} (\errnorms +\errtom + \epsilon'+ \errkappa)$ where $\epsilon'= \sqrt{ \lambda \beta^{2} + (1-\lambda) 4\eta}$ is a dataset dependent parameter computed in Claim \ref{low-rank}. We can set $\errkappa = \errnorms = \errtom = \epsilon'/3$ to obtain 
$\epsilon_{centroid} = 2\sqrt{\eta} \epsilon'$. 

The definition of the $\delta$-$k$-means update rule requires that $\epsilon_{centroid} \leq \delta/2$. Further, Claim \ref{distcentroid} shows that if the 
error $\delta$ in the assignment step satsifies $\delta \leq \xi^{2} - 2\sqrt{\eta} \beta$, then the $\delta$-$k$-means algorithm finds the corrects clusters. 
By Definition  \ref{wcdataset} of a well-clusterable dataset, we can find a suitable constant $\delta$
 satisfying both these constraints, namely satisfying
 $$4\sqrt{\eta} \sqrt{ \lambda \beta^{2} + (1-\lambda) 4\eta} < \delta <  \xi^{2} - 2\sqrt{\eta} \beta.$$

Substituting the values $\mu(V) = O(\sqrt{k})$ from Claim \ref{mu}, $\kappa(V) = O(\frac{1}{\epsilon_{\tau}})$ and $\errkappa = \errnorms = \errtom = \epsilon'/3 = O(\sqrt{\eta}/\delta)$ in the running time for the general $q$-means algorithm, we obtain that the running time for the $q$-means algorithm on a well-clusterable dataset is 
$\widetilde{O}\left( k^2 d \frac{\eta^{2.5}}{\delta^3} + k^{2.5} \frac{\eta^2}{\delta^3} \right)$ per iteration.

\end{proof}
%\noindent If a dataset is well-clusterable not only we have a good clustering assignment, but we can also choose a constant value for $\delta$ for which the $\delta$-$k$-means algorithm is able to find the clusters. In the case of a dataset satisfying the condition that the minimum norm is one of Definition \ref{datasetassumption}, the parameter $\eta/\delta$ can be interpreted as a relative error. In the next Section, we will try to get an estimate of these parameters, and conclude that for a simple dataset, are small enough.

Let us make some concluding remarks regarding the running time of $q$-means. For dataset where the number of points is much bigger compared to the other parameters, the running time for the $q$-means algorithm is an improvement compared to the classical $k$-means algorithm. For instance, for most problems in data analysis, $k$ is eventually small ($<100$). The number of features $d\leq N$ in most situations, and it can eventually be reduced by applying a quantum dimensionality reduction algorithm first (which have running time poly-logarithmic in $d$). To sum up, $q$-means has the same output as the classical $\delta$-$k$-means algorithm (which approximates k-means), it conserves the same number of iterations, but has a running time only poly-logarithmic in $\n$, giving an exponential speedup with respect to the size of the dataset.

\section{Simulations on real data}\label{experiment_section}

We would like to assert the capability of the quantum algorithm to provide accurate classification results, by simulations on a number of datasets. However, since neither quantum simulators nor quantum computers large enough to test $q$-means are available currently, we tested the equivalent classical implementation of $\delta$-$k$-means. 
For implementing the $\delta$-$k$-means, we changed the assignment step of the $k$-means algorithm to select a random centroid among those that are $\delta$-close to the closest centroid and added $\delta/2$ error to the updated clusters. 

We benchmarked our $q$-means algorithm on two datasets: a synthetic dataset of gaussian clusters, and the well known MNIST dataset of handwritten digits. To measure and compare the accuracy of our clustering algorithm, we ran the $k$-means and the $\delta$-$k$-means algorithms for different values of $\delta$ on a training dataset and then we compared the accuracy of the classification on a test set, containing data points on which the algorithms have not been trained, using a number of widely-used performance measures.

\subsection{Gaussian clusters dataset}
\begin{figure} 
\centering
\includegraphics[width=103mm, height=68mm] {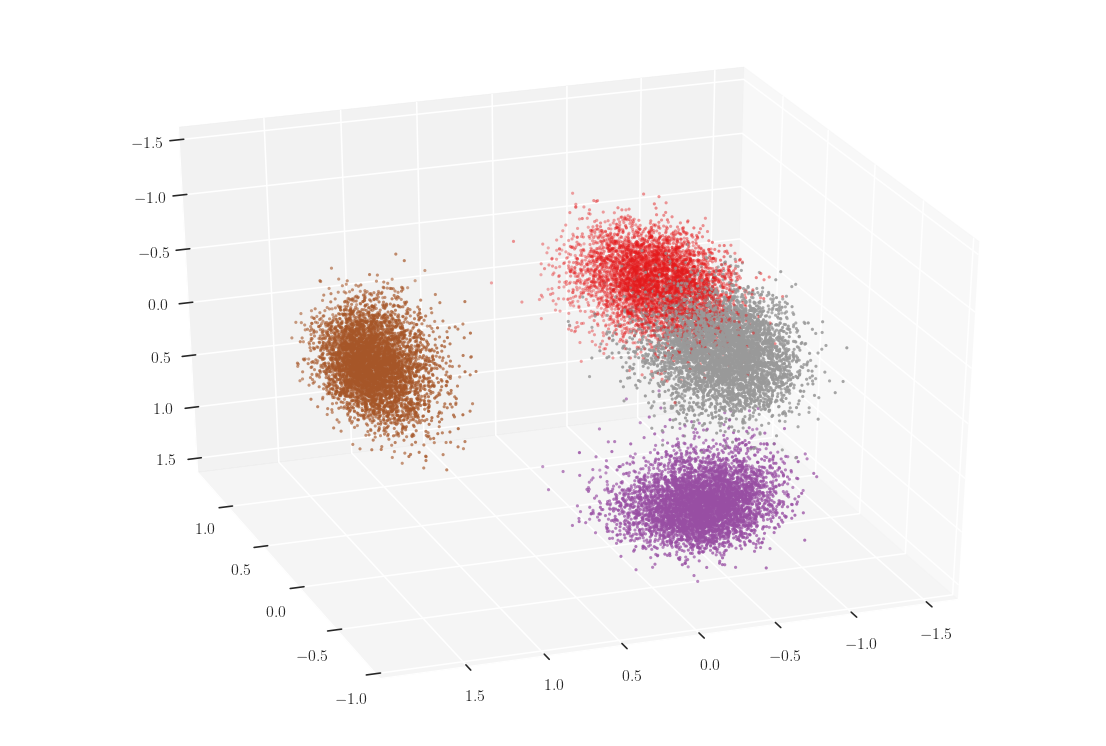} 
\captionsetup{justification=raggedright, margin=1cm}
\caption{Representation of $4$ Gaussian clusters of $10$ dimensions in a 3D space spanned by the first three PCA dimensions.}\label{gaussian-cluster-1}
\end{figure}

We describe numerical simulations of the $\delta$-$k$-means algorithm on a synthetic dataset made of several clusters formed by random gaussian distributions. These clusters are naturally well suited for clustering by construction, close to what we defined to be a well-clusterable dataset in Definition \ref{wcdataset} of Section \ref{datasetassumption}. Doing so, we can start by comparing $k$-means and $\delta$-$k$-means algorithms on high accuracy results, even though this may not be the case on real-world datasets. Without loss of generality, we preprocessed the data so that the minimum norm in the dataset is $1$, in which case $\eta = 4$. This is why we defined $\eta$ as a maximum instead of the ratio of the maximum over the minimum which is really the interesting quantity. Note that the running time basically depends on the ratio $\eta/\delta$.
We present a simulation where $20.000$ points in a feature space of dimension $10$ form $4$ Gaussian clusters with standard deviation $2.5$, that we can see in Figure \ref{gaussian-cluster-1}.  The condition number of dataset is calculated to be $5.06$. We ran $k$-means and $\delta$-$k$-means for $7$ different values of $\delta$ to understand when the $\delta$-$k$-means becomes less accurate.

\begin{figure}  [H] 
\centering
\includegraphics[width=102mm, height=68mm] {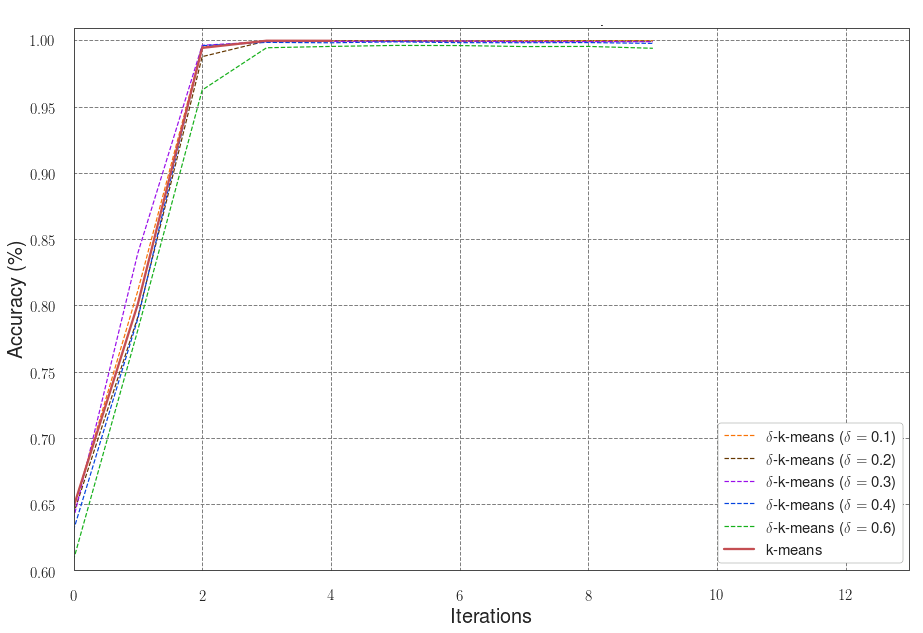} 
\captionsetup{justification=raggedright, margin=1cm}
\caption{Accuracy evolution during $k$-means and $\delta$-$k$-means on well-clusterable Gaussians for $5$ values of $\delta$. All versions converged to a 100\% accuracy in few steps.}\label{results-gaussian-cluster-1}
\end{figure}

In Figure \ref{results-gaussian-cluster-1} we can see that until $\eta/\delta = 3$ (for $\delta=1.2$), the $\delta$-$k$-means algorithm converges on this dataset. We can now make some remarks about the impact of $\delta$ on the efficiency. It seems natural that for small values of $\delta$ both algorithms are equivalent. For higher values of $\delta$, we observed a late start in the evolution of the accuracy, witnessing random assignments for points on the clusters' boundaries. However, the accuracy still reaches $100$\% in a few more steps. The increase in the number of steps is a tradeoff with the parameter $\eta/\delta$. 

\subsection{MNIST}

The MNIST dataset is composed of $60.000$ handwritten digits as images of 28x28 pixels (784 dimensions). From this raw data we first performed some dimensionality reduction processing, then we normalized the data such that the minimum norm is one. Note that, if we were doing $q$-means with a quantum computer, we could use efficient quantum procedures equivalent to Linear Discriminant Analysis, such as \cite{KL18}, or other quantum dimensionality reduction algorithms like \cite{lloyd2014quantum, cong2015quantum}. 

\begin{figure} [H]
\centering
\includegraphics[width=101mm, height=69mm] {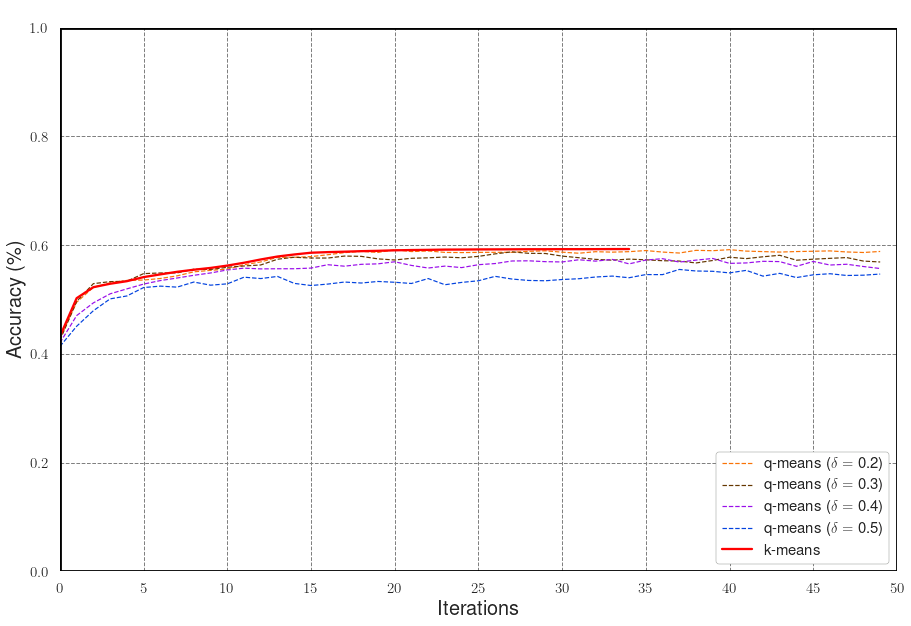} 
\captionsetup{justification=raggedright, margin=0.5cm}
\caption{Accuracy evolution on the MNIST dataset under $k$-means and $\delta$-$k$-means for $4$ different values of $\delta$. Data has been preprocessed by a PCA to 40 dimensions. All versions converge in the same number of steps, with a drop in the accuracy while $\delta$ increases. The apparent difference in the number of steps until convergence is just due to the stopping condition for $k$-means and $\delta$-$k$-means.} \label{mnist-results-1} 
\end{figure}

As preprocessing of the data, we first performed a Principal Component Analysis (PCA), retaining data projected in a subspace of dimension 40. After normalization, the value of $\eta$ was 8.25 (maximum norm of 2.87), and the condition number was 4.53. Figure \ref{mnist-results-1} represents the evolution of the accuracy during the $k$-means and $\delta$-$k$-means for $4$ different values of $\delta$. In this numerical experiment, we can see that for values of the parameter $\eta/\delta$ of order 20, both $k$-means and $\delta$-$k$-means reached a similar, yet low accuracy in the classification in the same number of steps. It is important to notice that the MNIST dataset, without other preprocessing than dimensionality reduction, is known not to be well-clusterable under the $k$-means algorithm.

On top of the accuracy measure (ACC), we also evaluated the performance of $q$-means against many other metrics, reported in Table \ref{tablecomparison} and \ref{tablecomparison2}. More detailed information about these metrics can be found in \cite{scklearncompare, friedman2001elements}. We introduce a specific measure of error, the Root Mean Square Error of Centroids (RMSEC), which a direct comparison between the centroids predicted by the k-means algorithm and the ones predicted by the $\delta$-$k$-means. It is a way to know how far the centroids are predicted. Note that this metric can only be applied to the training set. For all these measures, except RMSEC, a bigger value is better. Our simulations show that $\delta$-$k$-means, and thus the $q$-means, even for values of $\delta$ (between $0.2-0.5$) achieves similar performance to $k$-means, and in most cases the difference is on the third decimal point.

\begin{table}[H]
\centering
\begin{tabular}{|c|c|c|c|c|c|c|c|c|}
\hline
Algo                                  & Dataset & ACC   & HOM   & COMP  & V-M   & AMI   & ARI   & RMSEC \\ \hline
\multirow{2}{*}{k-means}              & Train   & 0.582 & 0.488 & 0.523 & 0.505 & 0.389 & 0.488 & 0     \\ \cline{2-9} 
                                      & Test    & 0.592 & 0.500 & 0.535 & 0.517 & 0.404 & 0.499 & -     \\ \hline
\multirow{2}{*}{$\delta$-$k$-means, $\delta=0.2$} & Train   & 0.580 & 0.488 & 0.523 & 0.505 & 0.387 & 0.488 & 0.009 \\ \cline{2-9} 
                                      & Test    & 0.591 & 0.499 & 0.535 & 0.516 & 0.404 & 0.498 & -     \\ \hline
\multirow{2}{*}{$\delta$-$k$-means, $\delta=0.3$} & Train   & 0.577 & 0.481 & 0.517 & 0.498 & 0.379 & 0.481 & 0.019 \\ \cline{2-9} 
                                      & Test    & 0.589 & 0.494 & 0.530 & 0.511 & 0.396 & 0.493 & -     \\ \hline
\multirow{2}{*}{$\delta$-$k$-means, $\delta=0.4$} & Train   & 0.573 & 0.464 & 0.526 & 0.493 & 0.377 & 0.464 & 0.020 \\ \cline{2-9} 
                                      & Test    & 0.585 & 0.492 & 0.527 & 0.509 & 0.394 & 0.491 & -     \\ \hline
\multirow{2}{*}{$\delta$-$k$-means, $\delta=0.5$} & Train   & 0.573 & 0.459 & 0.522 & 0.488 & 0.371 & 0.459 & 0.034 \\ \cline{2-9} 
                                      & Test    & 0.584 & 0.487 & 0.523 & 0.505 & 0.389 & 0.487 & -     \\ \hline
\end{tabular}
\caption{A sample of results collected from the same experiments as in Figure \ref{mnist-results-1}. Different metrics are presented for the train set and the test set. ACC: accuracy. HOM: homogeneity. COMP: completeness. V-M: v-measure. AMI: adjusted mutual information. ARI: adjusted rand index. RMSEC: Root Mean Square Error of Centroids.} 
\label{tablecomparison}
\end{table}

These experiments have been repeated several times and each of them presented a similar behavior despite the random initialization of the centroids. 

\begin{figure}[htbp]
\centering
\begin{minipage}{0.3\textwidth}
  \centering
\includegraphics[width=50mm, height=35mm]{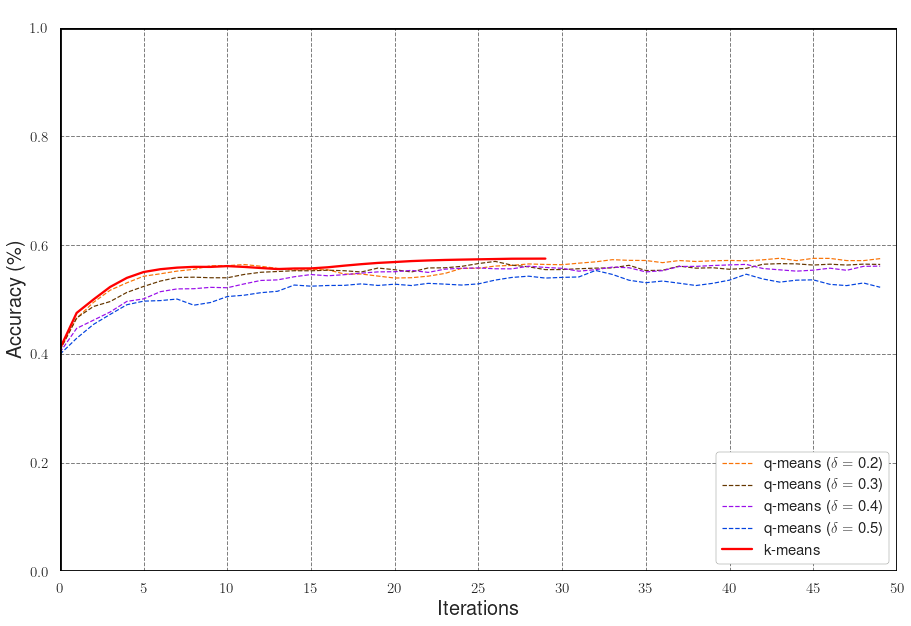}
\subcaption[second caption.]{}\label{fig:1a}
\end{minipage}%
\begin{minipage}{0.3\textwidth}
  \centering
\includegraphics[width=50mm, height=35mm]{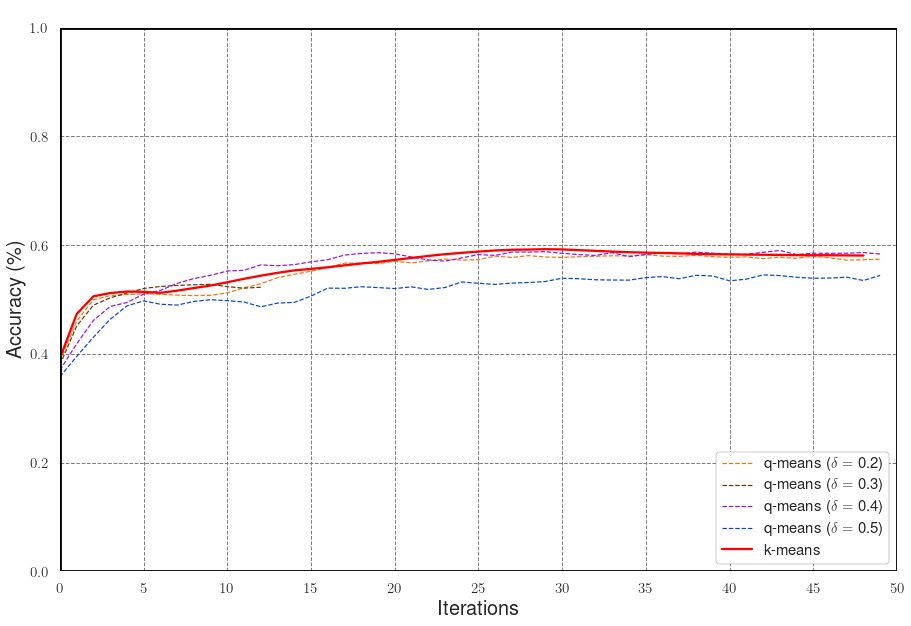}
\subcaption[third caption.]{}\label{fig:1b}
\end{minipage}
\begin{minipage}{0.3\textwidth}
  \centering
\includegraphics[width=50mm, height=35mm]{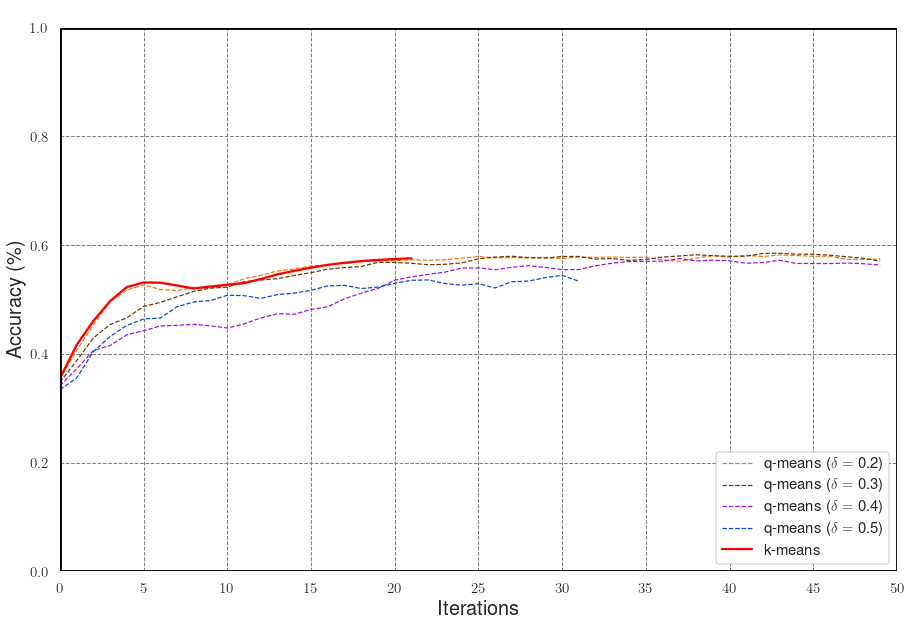}
\subcaption[third caption.]{}\label{fig:1c}
\end{minipage}
\captionsetup{justification=raggedright, margin=1cm}
\caption{Three accuracy evolutions on the MNIST dataset under $k$-means and $\delta$-$k$-means for $4$ different values of $\delta$. Each different behavior is due to the random initialization of the centroids} \label{fig:1}
\end{figure}

\begin{figure} [H]
\centering
\includegraphics[width=101mm, height=69mm] {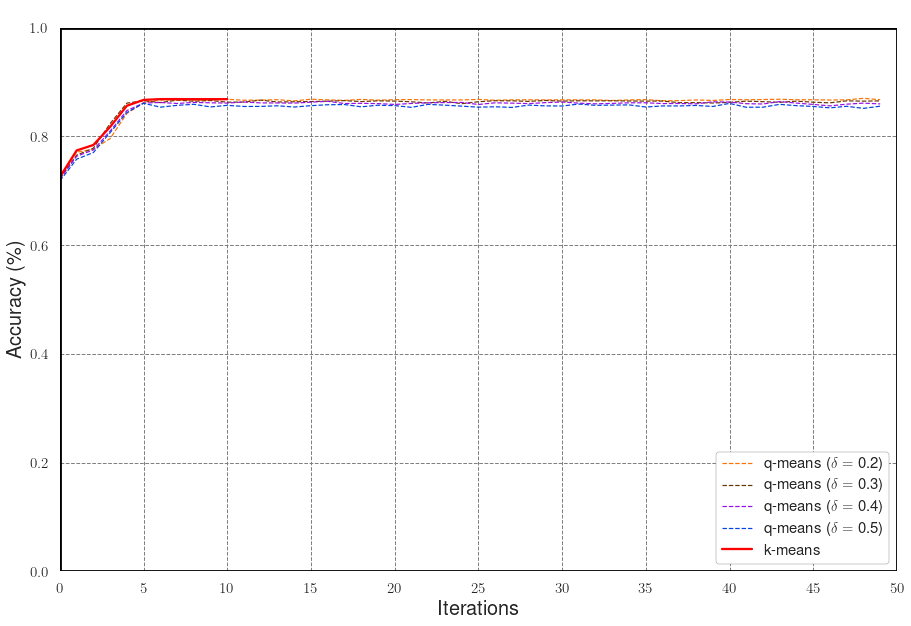} 
\captionsetup{justification=raggedright, margin=0.5cm}
\caption{Accuracy evolution on the MNIST dataset under $k$-means and $\delta$-$k$-means for $4$ different values of $\delta$. Data has been preprocessed to 9 dimensions with a LDA reduction. All versions of  $\delta$-$k$-means converge to the same accuracy than $k$-means in the same number of steps.} \label{mnist-results-2} 
\end{figure}

Finally, we present a last experiment with the MNIST dataset with a different data preprocessing. In order to reach higher accuracy in the clustering, we replace the previous dimensionality reduction by a Linear Discriminant Analysis (LDA). Note that a LDA is a supervised process that uses the labels (here, the digits) to project points in a well chosen lower dimensional subspace. Thus this preprocessing cannot be applied in practice in unsupervised machine learning. However, for the sake of benchmarking, by doing so $k$-means is able to reach a 87\% accuracy, therefore it allows us to compare $k$-means and $\delta$-$k$-means on a real and almost well-clusterable dataset. In the following, the MNIST dataset is reduced to 9 dimensions. The results in Figure \ref{mnist-results-2} show that $\delta$-$k$-means converges to the same accuracy than $k$-means even for values of $\eta/\delta$ down to $16$. In some other cases, $\delta$-$k$-means shows a faster convergence, due to random fluctuations that can help escape faster from a temporary equilibrium of the clusters.

\begin{table}[H]
\centering
\begin{tabular}{|c|c|c|c|c|c|c|c|c|}
\hline
Algo                                   & Dataset & ACC   & HOM   & COMP  & V-M   & AMI   & ARI   & RMSEC \\ \hline
\multirow{2}{*}{k-means}               & Train   & 0.868 & 0.736 & 0.737 & 0.737 & 0.735 & 0.736 & 0     \\ \cline{2-9} 
                                       & Test    & 0.891 & 0.772 & 0.773 & 0.773 & 0.776 & 0.771 & -     \\ \hline
\multirow{2}{*}{q-means, $\delta=0.2$} & Train   & 0.868 & 0.737 & 0.738 & 0.738 & 0.736 & 0.737 & 0.031 \\ \cline{2-9} 
                                       & Test    & 0.891 & 0.774 & 0.775 & 0.775 & 0.777 & 0.774 & -     \\ \hline
\multirow{2}{*}{q-means, $\delta=0.3$} & Train   & 0.869 & 0.737 & 0.739 & 0.738 & 0.736 & 0.737 & 0.049 \\ \cline{2-9} 
                                       & Test    & 0.890 & 0.772 & 0.774 & 0.773 & 0.775 & 0.772 & -     \\ \hline
\multirow{2}{*}{q-means, $\delta=0.4$} & Train   & 0.865 & 0.733 & 0.735 & 0.734 & 0.730 & 0.733 & 0.064  \\ \cline{2-9} 
                                       & Test    & 0.889 & 0.770 & 0.771 & 0.770 & 0.773 & 0.769 & -     \\ \hline
\multirow{2}{*}{q-means, $\delta=0.5$} & Train   & 0.866 & 0.733 & 0.735 & 0.734 & 0.731 & 0.733 & 0.079 \\ \cline{2-9} 
                                       & Test    & 0.884 & 0.764 & 0.766 & 0.765 & 0.764 & 0.764 & -     \\ \hline
\end{tabular}
\caption{A sample of results collected from the same experiments as in Figure \ref{mnist-results-2}. Different metrics are presented for the train set and the test set. ACC: accuracy. HOM: homogeneity. COMP: completeness. V-M: v-measure. AMI: adjusted mutual information. ARI: adjusted rand index. RMSEC: Root Mean Square Error of Centroids.} 
\label{tablecomparison2}
\end{table}

Let us remark, that the values of $\eta/\delta$ in our experiment remained between 3 and 20. Moreover, the parameter $\eta$, which is the maximum square norm of the points, provides a worst case guarantee for the algorithm, while one can expect that the running time in practice will scale with the average square norm of the points. For the MNIST dataset after PCA, this value is 2.65 whereas $\eta = 8.3$. 

In conclusion, our simulations show that the convergence of $\delta$-$k$-means is almost the same as the regular $k$-means algorithms for large enough values of $\delta$. This provides evidence that the $q$-means algorithm will have as good performance as the classical $k$-means, and its running time will be significantly lower for large datasets.

\bibliographystyle{IEEEtran} 

\bibliography{qmeans} 

\end{document}